\documentclass[12pt,twoside]{article}
\usepackage{correl}
\def\TheTitle{Why Calculate in Infinite Dimensions?} %\index{infinite dimensions}}
\def\TheHeading{Why $d\,{=}\,\infty$ dimensions?}
\def\TheAuthor{Dieter Vollhardt}
\def\TheInstitute{Center for Electronic Correlations and Magnetism}
\def\TheAddress{University of Augsburg\vspace{-18ex}}
\def\TheChapter{1}

\newcommand{\be}{\begin{equation}}
\newcommand{\beq}{\begin{eqnarray}}
\newcommand{\eeq}{\end{eqnarray}}
\newcommand{\ee}{\end{equation}}

\newcommand{\Hop}{\hat{H}}

\renewcommand{\epsilon}{\varepsilon}
\renewcommand{\rho}{\varrho}
\renewcommand{\phi}{\varphi}
\renewcommand{\theta}{\vartheta}
\newcommand{\texorpdfstring}[2]{#1}

\begin{document}
\MakeTitle

\section{Dimensions in physics: From zero to infinity}

In physics the term \emph{dimension}\footnote{\textbf{Dimension}: from Latin \emph{dimensio} ``a measuring'', noun from past participle stem of \emph{dimetiri} ``to measure'' \cite{OxfordDict}.} has a double meaning. On the one hand, it denotes the number of coordinates required to specify the location of a point within an object (or, more generally, a mathematical space). In particular, the dimension $d$ of a point is $d=0$, of a line $d=1$, of a surface $d=2$, and of a solid body $d=3$.
%The concept of \emph{spatial} dimensions plays a particularly important role in classical Newtonian mechanics, where time and space are distinct categories.
On the other hand, Maxwell introduced the dimension of a physical quantity, which is expressed in powers of fundamental units of mass, length, time, etc.~\cite{Roche+Bramwell}.
Thereby it became possible to analyze the relationship between different physical quantities (``dimensional analysis''), which was later put on a firm theoretical basis by renormalization group theory.

In this lecture the term dimension refers to \emph{spatial} dimensions.

\subsection{Integer and continuous spatial dimensions}

\subsubsection{Integer dimensions}
In theoretical physics the dimensionality of a microscopic model with finite-range interactions (or, for quantum-mechanical models in tight-binding approximation with finite-range hopping amplitude) in all directions, corresponds to the dimension of the space considered. This is particularly evident in the case of models on regular lattices, such as the hypercubic lattice, which is a straightforward generalization of the simple cubic lattice ($d=3$) to lower ($d=0,1,2$) or higher ($d=4,5,6, ..., \infty$) dimensions. The usefulness of investigations of models in dimensions higher than the ``real'' dimension $d=3$ is not self-evident. In their 1964 paper on the Ising model on general $d$-dimensional hypercubic lattices Fisher and Gaunt \cite{Fisher+Gaunt} found it necessary to write:
\begin{list}{}{\leftmargin4ex\rightmargin\leftmargin}
\item\emph{Of course the behavior of model physical systems in four or more space-like dimensions is not directly relevant to comparison with experiment! We can hope, however, to gain theoretical insight into the general mechanism and nature of phase transitions.}
\end{list}
%%e c_N: number of sites that can be reached with at most N steps
%%e hypercubic: number of lattice points in hypersphere of radius N*a: e.g. d=3: \ln c_N = \ln 4\pi/3 + 3\ln N  (up to discretization...)
%%e volume of hypersphere in d dimensions: V_d = \pi^{d/2}/\Gamma(d/2+1) (also used at the end of Sec. 2.3)
The dimensionality of a lattice may be deduced from the total number of sites which can be visited in a walk of $N$ steps away from a given site. The higher the dimension $d$, the more sites there are. For regular $d$-dimensional lattices and large $N$ the number of sites in such a ``sphere'' of radius $N$ is proportional to $N^d$. This can be formalized as follows \cite{Baxter}: if $c_N$ denotes the number of sites which are $N$ steps from a given site, one finds for hypercubic lattices
\begin{equation}
\lim_{N\rightarrow \infty} \frac{\ln c_N}{\ln N}= d,
\label{def-dim}
\end{equation}
where $d$ is the dimensionality of the lattice. This relation, which also holds for regular two- and three-dimensional lattices, may therefore be used as a definition of the dimensionality $d$ of a lattice-like structure.
More specifically, for regular lattices both a dimension $d$ and a \emph{coordination number} $Z$ (the number of nearest neighbors of a site) can be defined; $Z$ is then determined by the dimension $d$ and the lattice structure. In particular, for a hypercubic lattice
one has $Z=2d$. But there exist lattice-like structures without a relation between $d$ and $Z$. A famous example is the Bethe lattice (which is not a lattice but an infinite, loop-free graph (``Cayley tree'') without lattice periodicity), where each site has the same number of nearest neighbors $Z$. In this case the number of sites which are  $N$ steps away from a given site is $c_N=Z\big((Z{-}1)^N-1\big)/(Z{-}2)$ for $Z\geq 2$, i.e., it increases exponentially with $N$\!.\, This is a faster increase than $N^d$ for any finite $d$. According to the definition \eqref{def-dim} the Bethe lattice is therefore effectively infinite-dimensional.
%%e Bethe lattice: no lattice periodicity but all sites are equivalent
%%e number of n-th nearest neighbors: Z_n=Z*(Z-1)^{n-1}; special cases: Z=1: dimer, Z=2: chain
%%e number of sites that can be reached with at most N steps: sum_{n=0}^N Z_n = Z* ((Z-1)^N-1)((Z-1)-1)

\subsubsection{Continuous dimensions}

Historically, the concept of spatial dimensions implied that the dimension is a fixed, integer number, e.g., $d=3$. But, both from a mathematical and physical point of view this is an unnecessary restriction.\footnote{For example, there exists a class of fractal objects \cite{Mandelbrot} (``fractals''), whose defining property is their scale invariance or self-similarity. A quantitative measure of a fractal is its dimension (``fractal dimension'') which is, in general, not an integer. For example, a rocky coast line has a fractal dimension between $d=1$ and $d=2$. It is even possible to design and characterize electrons in fractal geometries \cite{Kempkes2019}. A discussion of the notion of dimension in a more general context can be found in ref.~\cite{Bhatta2017}.} In particular, with the formulation of the renormalization group theory \cite{RG} it became apparent that it is useful to regard the spatial dimension $d$  as a continuous parameter and to
analytically continue $d$-dimensional integrals from integer to continuous values of $d$. This led to the ``$\epsilon$-expansion'', with $\epsilon=4-d$, which for continuous values of $d$ close to the critical dimension $d=4$ of the Ginzburg-Landau theory can be arbitrarily small ($\epsilon\ll1$). Expansions in $\epsilon$ then make it possible to perform explicit perturbative calculations around $d=4$, as expressed by the striking title ``Critical Exponents in $3.99$ Dimensions'' of a paper by Wilson and Fisher \cite{Wilson+Fisher}.

\subsection{Simplifications arising in infinite dimensions}\index{infinite dimensions}
\label{Infinite-dimensions}

The mathematical equations describing classical or quantum-mechanical many-body systems can almost never be solved exactly in dimension $d=3$. In many problems there does not even exist a small physical parameter (``control parameter''), in which an expansion could be performed (if such an expansion is tractable at all).
One way out is to study problems in $d{=}1$, where mathematical solutions are more readily available. Such investigations have indeed led to a wealth of insight. However,  one-dimensional systems are quite special and do not describe behavior which is characteristic for systems in $d>2$, e.g., thermal phase transitions. An alternative
is to go in the opposite direction. Indeed, the coordination number of a three-dimensional lattice can already be quite large, e.g., $Z = 6$ for a simple cubic lattice, $Z = 8$ for a bcc lattice and $Z  = 12$ for an
fcc-lattice. It is then interesting to
check whether the limit $Z \to  \infty$ leads to some simplifications.
Such investigations do not go far back in time. In fact, $Z$ was originally regarded as a measure of the \emph{range} of the interaction between spins in the Ising model, and thus of the number of spins in the range of the interaction \cite{Brout60}.
In this case
the limit $Z\rightarrow \infty $ describes an infinitely long-ranged interaction.\footnote{This limit is even useful for one-dimensional models. For example, for a solvable one-dimensional particle model it was shown \cite{Kac63-64} that in this limit the equation of state reduces to the van der Waals equation.} Since a particle or spin then interacts with infinitely many other particles or spins (which are all ``neighbors'', i.e., the system has `infinite connectivity''`), this limit was originally referred to as ``limit of high density'' \cite{Brout60} and later as ``limit of infinite dimensions''  \cite{Fisher+Gaunt}.
Starting with Fisher and Gaunt \cite{Fisher+Gaunt} the Ising model\index{Ising model} and other classical models were investigated on general $d$-dimensional hypercubic lattices. Today $Z$ denotes the coordination number, i.e., the number of nearest neighbors.

\subsubsection{Infinite dimensions and mean-field behavior}\index{mean-field theory}

In the statistical theory of classical and quantum-mechanical systems mean-field theories play an important role since they provide an approximate understanding of the properties of a model. While in the full many-body model a particle or spin experiences a complicated, fluctuating field generated by the other particles or spins, a mean-field theory introduces an average field (``mean field'') instead. Usually the full interaction problem then reduces to an effective, single-particle problem, which is described by a self-consistent field theory.

A mean-field theory can often be derived systematically by increasing the range of an interaction (e.g., the coupling between spins) or the size of a parameter (e.g., the spin or orbital degeneracy $N$\!,\, the spatial dimension $d$, or the coordination number  $Z$) to infinity. In particular, the mean-field theory of the Ising model (the so called ``molecular'' or ``Weiss'' mean-field theory, which can be derived by the Bragg-Williams approximation \cite{Huang}), becomes exact in the limit $d \to \infty$; see section \ref{sec:ising}.\footnote{Due to the simplicity of the Ising model the limit of infinitely long-ranged spin coupling $J$ and of infinite dimensions $d$ both yield the same mean-field theory. However, for more complicated models, in particular quantum models with itinerant degrees of freedom, this will in general not be the case.}
 This also answers the question serving as the title of this lecture \emph{Why calculate in infinite dimensions?} Namely,  equations
 which depend on the dimension $d$ and which are too complicated to be solved in $d=3$, often simplify in the limit $d \to  \infty$ to such a degree that they can be solved exactly.\footnote{To avoid trivial results or divergencies when taking the limit $d\rightarrow \infty$, an appropriate scaling of parameters or coupling constants is necessary, as will be discussed later.}
 At best this leads to an approximate solution which retains characteristic features of the problem in $d<\infty$ and  provides insights into the (unknown) solution in $d=3$. Several examples will be discussed in this lecture.
To this end I will address  mainly many-body systems (classical and quantum) of condensed matter physics and the simplifications arising in $d=\infty$. However, to demonstrate the usefulness of this approach I first discuss the solution of a famous quantum-mechanical \emph{two}-body problem, the hydrogen atom, in the limit $d\rightarrow \infty$.
%%e one-body ? (nucleus considered fixed...)

\subsection[Example: Derivation of Bohr's atomic model from the Schr{\"o}dinger equation]{Example:\newline Derivation of Bohr's atomic model from the Schr{\"o}dinger equation}\index{Bohr's atomic model}

Bohr's model of the hydrogen atom is semiclassical: it postulates that in the ground state the electron moves in a circular orbit around the proton, like a planet around the sun in classical mechanics, and assumes in addition that the angular momentum and the energy of the electron are quantized. Although this model was extremely useful in the pre-history of quantum mechanics\footnote{Even today the Bohr model plays a useful role, e.g., in the description of highly excited Rydberg atoms \cite{Gallagher} and cavity quantum electrodynamics \cite{Haroche}.} its connection to the ``proper'' quantum mechanics was unclear for a long time. Today we understand that the main features of Bohr's model may be derived by solving the hydrogen atom with the Schr{\"o}dinger equation in infinite dimensions. This approach was introduced by Witten  \cite{Witten} to demonstrate the usefulness of expansions in the inverse of some large parameter for deriving approximate solutions of  hard problems such as quantum chromodynamics. The results for the hydrogen problem (which is, of course, exactly solvable in $d=3$) showed that even atomic physics greatly simplifies as one goes from $d=3$ to $d=\infty$, and that expansions in $1/d$ provide valuable approximations for problems which are no longer exactly solvable \cite{Witten,Scully}.

Following ref.\cite{Scully} the radial Schr{\"o}dinger equation in $d$-dimensional spherical coordinates for the electron of a hydrogen atom with infinite proton mass and orbital angular momentum $l$ is given by
\begin{equation}
\left [ - \frac{1}{2}
\left (
{\frac{\mathrm{d}^2}{\mathrm{d}\rho^2}} +
\frac{d{-}1}{\rho}\frac{\mathrm{d}}{\mathrm{d}\rho}
\right ) + \frac{l(l{+}d{-}2)}{2\rho^2} - \frac{1}{\rho}
\right ] R = \epsilon R,
\label{radial-Schrod-eq}
\end{equation}
where
%$R(\rho)$ is the radial wavefunction and
we used atomic units, i.e., $\rho=r\!/a_0$ is the radial distance in units of the Bohr radius
$a_0 = {\hbar}^{2}\!/(e^{2} m)$
and $\epsilon=E/(e^2/a_0)$ is
the energy of the electron in units of Hartree.\footnote{Note that the Coulomb potential of the hydrogen atom is that in $d=3$, not in general $d$ (in which case it would have a $1/\rho^{d-2}$ dependence for $d>2$).}
To eliminate the first-order derivative in \eqref{radial-Schrod-eq} we write
the radial wavefunction
$R$ as $R(\rho)=u(\rho){\rho}^{-(d-1)/2}$, whereby \eqref{radial-Schrod-eq}
reduces to
\begin{equation}
\left ( - \frac{1}{2}
{\frac{\mathrm{d}^2}{\mathrm{d}\rho^2}} +
\frac{\Lambda(\Lambda{+}1)}{2\rho^2} - \frac{1}{\rho}
\right )u = \epsilon u.
\label{radial-Schrod-eq-scaled}
\end{equation}
The $d$-dependence now enters  only in the centrifugal potential, where $\Lambda=l+(d{-}3)/2$ replaces the usual orbital angular momentum. As a consequence, the main quantum number $n$ in $d=3$ becomes $n+(d{-}3)/2$ in $d\geq3$.
Since the form of \eqref{radial-Schrod-eq-scaled}, which determines
$u(\rho)$ in $d\geq3$, is the same as that in $d=3$, all properties of the $d$-dimensional hydrogen atom can be related to those of the well-known solution in $d=3$ (in the following we discuss only the ground state, $l=0$). In particular, one finds that
the radial probability distribution
%${\lvert u \rvert}^2$
is maximal at the distance
$r_{\rm max}=(\frac{d-1}{2})^2 a_0$.  As $d$ increases the width of the electron distribution \emph{decreases}, i.e., the electron is confined to a thinner and thinner spherical shell
of radius $r_{\rm max}$
around the nucleus. In addition, the factor $(\sin \theta)^{d-2}$ in the $d$-dimensional volume element restricts the polar angle  to $\theta \rightarrow \pi/2 $ for $d\rightarrow \infty$.  Hence for large $d$ the electronic probability distribution approaches the shape of a planar, circular orbit as described by the Bohr model!

To make sure that the centrifugal term in \eqref{radial-Schrod-eq-scaled}  remains finite in the limit $d\rightarrow \infty$ a dimensional rescaling of the radial coordinate and the energy is performed as $\mathcal{R}=(\frac{d-1}{2})^{-2} \rho$ and $\mathcal{E}=(\frac{d-1}{2})^{2}\epsilon$, respectively. This brings \eqref{radial-Schrod-eq-scaled} into the form
\begin{equation}
\left ( - \frac{2}{(d{-}1)^2}
{\frac{\mathrm{d}^2}{\mathrm{d}\mathcal{R}^2}} +
\frac{1}{2} \left .\frac{d{-}3}{d{-}1}\right . \frac{1}{\mathcal{R}^2} - \frac{1}{\mathcal{R}}
\right ) u = \mathcal{E} u.
\label{radial-Schrod-eq-scaled2}
\end{equation}

In $d=\infty$ the derivative in \eqref{radial-Schrod-eq-scaled2}, which originates from the kinetic energy of the electron, is suppressed and the equation reduces to the classical energy expression
\begin{equation}
\mathcal{E(\mathcal{R})}=
\frac{1}{2\mathcal{R}^2} - \frac{1}{\mathcal{R}}.
\label{Bohr-energy}
\end{equation}
This is indeed the energy of the hydrogen atom in  Bohr's model. It is minimal for
%determines the radial coordinate and energy of the electron 
%in $d=\infty$
%as 
$\mathcal{R}=1$, corresponding to $\mathcal{E_{\rm min}}=-1/2$, in agreement with the three-dimensional result. Hence in infinite dimensions the electron no longer orbits around the nucleus, but is located at the minimum of the effective potential. Quantum fluctuations are then completely quenched.\footnote{This does not violate the Heisenberg uncertainty relation since in the product of length and momentum the dimensional scaling factors cancel.} For that reason the electron does not emit radiation (photons) as had to be assumed by Bohr in 1913. This assumption is found to be justified in $d=\infty$.

The large-$d$ approach to the Schr{\"o}dinger equation outlined above can also be applied to more complicated problems of atomic physics, such as the helium atom or the hydrogen molecule. It turns out that perturbative calculations in the small parameter $1/d$ yield surprisingly accurate results for the two-electron bond \cite{Scully}.

As explained above, the solution of classical and quantum-mechanical many-body problem  obtained in the limit $d=\infty$ (``mean-field theory'') may provide important physical insights, which are not available otherwise. Since the mean-field theory of the classical Ising model is particularly simple and its validity in $d=\infty$ follows directly from the law of large numbers,
we start with this famous example, then discuss the simplifications arising in $d=\infty$ in other classical  model systems, and finally turn to interacting fermions and derive the dynamical mean-field theory (DMFT) of correlated electrons.

\section[Construction of classical mean-field theories in infinite dimensions]{Construction of classical mean-field theories\newline in infinite dimensions}\index{infinite dimensions}\index{mean-field theory}
\label{MFT}

\subsection{Ising model}\index{Ising model}\index{ferromagnetism}
\label{sec:ising}

In 1906 Weiss \cite{Weiss} introduced a model of magnetic domains, where he assumed that the alignment of the ``elementary magnets'' in each domain is caused by some sort of ``molecular field'', today referred to as ``Weiss mean field''. But how does this molecular field arise in the first place?
To answer this question and to explain ferromagnetism in three-dimensional solids from a truly microscopic point of view, Ising investigated a minimal microscopic model of interacting classical spins with a \emph{non-magnetic} interaction between neighboring elementary magnets \cite{Ising}, which had been proposed to him by his thesis advisor Lenz
in 1922.
The Hamiltonian function for the Ising model with coupling $J_{ij}$ between two spins at lattice sites $\bm{R}_{i}$, $\bm{R}_{j}$ is given by\footnote{This notation of the Ising model, as it is used today, was actually introduced by Pauli at a Solvay conference in 1930; a detailed discussion of the history of the Ising model and its many applications can be found in ref.~\cite{Ising-review}.}
\begin{equation}
 H  = - \frac{1}{2}
 \sum_{\bm{R}_{i}, \bm{R}_{j}}
\;J_{ij}\, S_i S_j.
\label{Ising-general}
\end{equation}
In particular, if the coupling is restricted to nearest-neighbor spins it takes the form
\begin{equation}
 H  = - \frac{1}{2}
J \sum_{\langle \bm{R}_{i}, \bm{R}_{j} \rangle}
\; S_i S_j,
\label{G11.1}
\end{equation}
where we assume ferromagnetic coupling ($J > 0$) and ${\langle \bm{R}_{i}, \bm{R}_{j} \rangle}$ indicates summation over all nearest-neighbor sites (the factor $1/2$ prevents double counting of sites).
This can also be written as
\begin{equation}
 H  = - \sum_{\bm{R}_{i}}\; h_i S_i,
\label{G11.1a}
\end{equation}
where now every spin $S_i$ interacts with a site-dependent, i.e., locally fluctuating, field
\begin{equation}
h_{i} =  J \sum_{\bm{R}_j}^{(i)} S_j
\label{G11.2ba} \\[10pt]
\end{equation}
generated by the coupling of the spin $S_i$ to its neighboring sites; here the superscript $(i)$ on the summation symbol indicates summation over the $Z$
nearest-neighbor sites of $\bm{R}_i$.

\subsubsection{Weiss mean field}\index{mean field}

In the Weiss mean-field theory
the interaction of a spin with its local field in \eqref{G11.1a} is decoupled (factorized),
i.e., $h_i$ is replaced by a mean field $h_{\rm MF}$, which leads to the mean-field Hamiltonian
\label{G11.2}
\begin{equation}
H^{\rm MF} =  - h_{\rm MF} \sum_{\bm{R}_i} \; S_i + E_{\rm shift}.
\label{G11.2a}
\end{equation}
Now a spin $S_i$ interacts only with a global  field
$h_{\rm MF} =  JZS$ (the ``molecular'' or ``Weiss'' field), where $S\equiv \langle S_i \rangle=(1/L) \sum_{i=0}^{L} S_i$ is the average value of $S_i$, $E_{\rm shift} = L J Z S{^2}/2$ is a constant energy shift, and $L$ is the number of lattice sites of the system.

\begin{figure}
\center
\includegraphics[width=0.9\textwidth,clip=true]{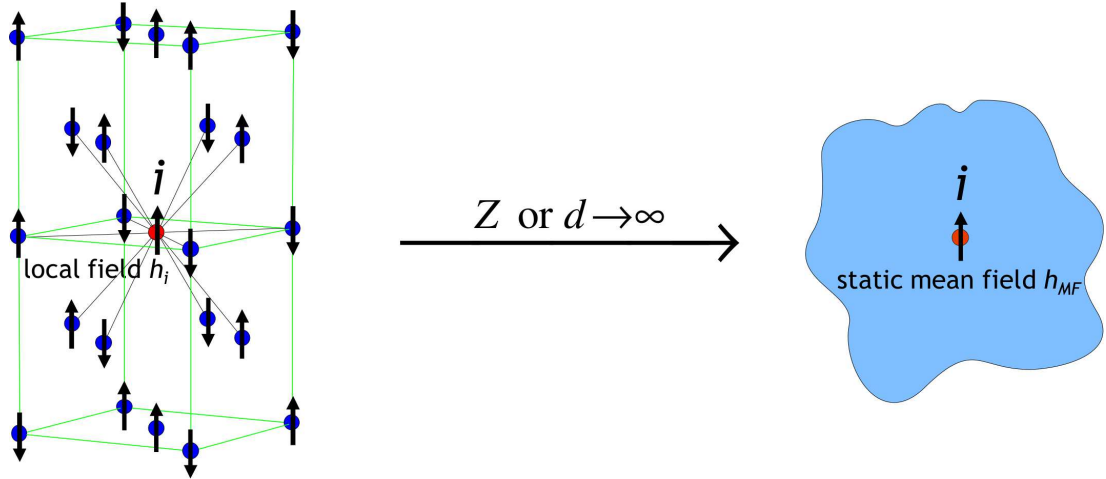}
\caption{Already in three dimensions ($d=3$) the coordination number $Z$ of a lattice can be quite large, as in the face-centered cubic lattice shown on the left, where $Z=12$. In the limit $Z \to \infty$, or equivalently $d \to \infty$, the Ising model effectively
reduces to a single-site problem where the local field $h_i$ is replaced
by a global (``molecular'')  mean field $h_{\rm MF}$.}
\label{Ising_MF}
\end{figure}

Next we show that in infinite dimensions or for infinite coordination number $Z$  this decoupling arises naturally.
First we have to rescale the coupling constant $J$ as
\begin{equation}
J =  \frac{J^*}{Z} \; , \; J^* = {\rm const},
\label{G11.3}
\end{equation}
so that $h_{\rm MF}$ and the energy (or the energy density in the thermodynamic limit) remain finite
in the limit $Z \to \infty$.
Writing $S_i=S+\delta S_i$, where $\delta S_i$ is the deviation of $S_i$ from its average $S$, \eqref{G11.2ba} becomes $h_{i} =  J^{*}(S+\Delta S_i)$, where
\begin{equation}
\Delta S_{i} =
\frac{1}{Z} \sum_{\bm{R}_j}^{(i)} \delta S_j
% \rightarrow \frac{1}{\sqrt{Z}}
\label{DeltaS}
\end{equation}
is the sum over the fluctuations  $\delta S_j$ of the $Z$ nearest-neighbor spins of $S_i$ per nearest neighbor. These fluctuations are assumed to be uncorrelated, i.e., random. The law of large numbers then implies that the sum increases only as $\sqrt{Z}$ for $Z \to \infty$, such that $\Delta S_{i}$ altogether \emph{decreases} as $1/\sqrt{Z}$ in this limit. As a consequence the local field
$h_{i}$ can indeed be replaced by its mean $h_{\rm MF}$ (central limit theorem). %
Hence the
Hamiltonian function \eqref{Ising-general} becomes purely local
\begin{equation}
H^{\rm MF}  = \sum_{\bm{R}_i} H_i + E_{\rm shift},
\label{G11.4}
\end{equation}
where $H_i =  - h_{\rm MF} S_i$. Thereby the problem reduces to an effective single-site problem
(see Fig.~\ref{Ising_MF}).
We note that  $S$ corresponds to the magnetization $m$ of the system ($S \equiv m$). In the paramagnetic phase, where $m=0$, the mean field $h_{\rm MF}$ vanishes; hence \eqref{G11.2a} and \eqref{G11.4} are only non-trivial in the presence of ferromagnetic order.
The magnetization $m$ is obtained from the partition function of the mean-field Hamiltonian \eqref{G11.2a} as  $m=\tanh(\beta h_{\rm MF})$ where
$\beta = 1/(k_B T)$ \cite{Huang}.
The self-consistency condition $h_{\rm MF}=JZS\equiv J{^*} m$ then yields the well-known self-consistent equation\index{self-consistent equation} for the magnetization $m$ as
\begin{equation}
m = \tanh \!\big(\beta J^{*} m\big).
\label{G11.6}
\end{equation}

The Weiss mean-field theory is seen to become exact in the limit of infinite coordination
number $Z$ or dimension $d$.
In this case $1/Z$ or $1/d$ serve as a small parameter which can  be used, in principle, to improve the
mean-field theory systematically (see section \ref{1-over-d-corrections}). This mean-field theory contains no
unphysical singularities, is  applicable for all values of the input
parameters (temperature and/or external magnetic field)
and is often viewed as a prototypical mean-field theory in statistical mechanics.

\subsection{Ising model with random coupling: The spin glass problem}\index{Ising model}\index{spin glass}
\label{Spin-glass}

The term \emph{spin glass} was introduced in the early 1970s to characterize the behavior of certain disordered magnetic systems, i.e., alloys of a non-magnetic material with a few percent of randomly distributed magnetic impurities, such as manganese in zinc oxide. At low temperatures a phase transition occurs, where the magnetic moments ``freeze'' into a state in which the spin orientations are not aligned, but randomly oriented. The disorder is due to the random local distribution of the spins and the fact that their coupling can be ferro- or antiferromagnetic. Theoretical investigations of this experimental finding started in 1975 with Edwards and Anderson \cite{Edwards-Anderson75} who proposed a classical Heisenberg model with random finite-range couplings $J_{ij}$, which can be positive or negative.  In particular, they constructed an order parameter for the spin glass phase and introduced the famous ``replica trick''. Namely, to compute the averaged free energy they considered $n$ copies (``replicas'') of the system, calculated the corresponding partition function $\mathcal{Z}^n$  to obtain $\ln \mathcal{Z}=\lim_{n\rightarrow 0} \big((\mathcal{Z}{^n}{-}1)/n\big)$ by taking the limit $n \to 0$, and then averaged  $\ln \mathcal{Z}$ over the randomness.
Later in the year Sherrington and Kirkpatrick \cite{Sherrington75} introduced a mean-field version of the Edwards-Anderson model by assuming Ising spins instead of Heisenberg spins, but with random \emph{infinite-range} spin couplings $J_{ij}$ in \eqref{Ising-general}, whereby all spins are equally coupled. For random $J_{ij}$ with zero mean
 ($\overline{J}_{ij} = 0$) but $\overline{J_{ij}^2} \neq 0$ the spin couplings need to be scaled as $J_{ij} \to J_{ij}^* /\sqrt{L}$, where $L$ is the number of lattice sites, to keep the energy density of the system finite in the thermodynamic limit $L \to \infty$. As mentioned in section \ref{Infinite-dimensions} the mean-field theory of the Ising model derived for infinite range-coupling is equivalent to that for finite-range coupling in infinite dimensions $d$ or for infinite coordination number $Z$. In the latter case, and for spins with random nearest-neighbor couplings $J_{ij}$, the scaling
\begin{equation}
J_{ij} = J_{ij}^* /\sqrt{Z}
\label{Spinglass-scaling}
\end{equation}
applies; in both cases the model has infinite connectivity. Thereby a mean-field investigation of the spin glass problem became possible which, however, resulted in a negative entropy at low temperatures \cite{Sherrington75}. This non-physical result was found to originate from the assumption of ``replica symmetry'' \cite{deAlmeida-Thouless78}. The question of how to break this symmetry (``replica symmetry breaking'') was answered by Parisi in 1979 \cite{Parisi79a+79b}, who realized that there must be an infinite number of order parameters in the spin glass phase. His analytic procedure led to a consistent and stable mean-field solution of the spin glass problem (for details see ref.~\cite{Billoire08+Parisi-PT-Dec2021}).\footnote{Parisi was awarded one half of the Nobel Prize in Physics 2021 ``for groundbreaking contributions to our understanding of complex systems''. The Nobel Committee highlighted the great influence which Parisi's concept of broken replica symmetry and his method of solution of the mean-field theory of the spin-glass problem had on the statistical physics of systems exhibiting multiple equilibria.
%, including its experimental observation, e.g. in the field of random lasers, were.
}

The mean-field theory of the spin-glass problem illustrates particularly well that, in spite of the simplifications introduced by a mean-field theory, the solution of this mean-field theory can still be extraordinarily complicated. This will become evident again in the context of the dynamical mean-field theory (DMFT) of correlated electrons (see section \ref{DMFT-equations}).

The limit of infinite dimensions and the methods of solution developed in the context of the spin-glass mean-field theory have also been very useful in the study of classical liquids and amorphous systems, such as glasses and granular matter, where calculations are notoriously difficult.  This made it possible to explore not only the thermodynamic properties but even the general dynamics of liquids and structural glasses, including the famous glass transition and the rheology of glasses, on a mean-field level \cite{Parisi-Zamponi-RMP2010+Manacorda2020,Parisi-book2020}.

\subsection{Hard-sphere fluid}\index{hard-sphere fluid}

As mentioned above, the limit of infinite spatial dimensions leads to theoretical simplifications not only in the study of lattice systems (in which case the coordination number $Z$ tends to infinity) but also in the case of systems in the continuum. This may be demonstrated by calculating the equation of state of a classical dilute hard-sphere fluid\footnote{A fluid is a substance which continuously deforms under tangential stress. Any liquid or gas is therefore a fluid (but not all fluids are liquids).}; here we follow ref.~\cite{Frisch85+87}.
While at low densities the virial series, i.e., the expansion of the equation of state in terms of the density, is a very useful technique to examine thermodynamic properties of a fluid in $d=3$, this approach breaks down at high densities. By contrast, these calculations are possible in infinite spatial dimensions\footnote{In the following $P$ is the pressure, $V$ the volume, $N$ the particle number, $n=N/V$ the number density, $\mu$ the chemical potential, and $m$ the mass of the hard sphere particles.} since in this limit the virial expansion $\beta PV=N\sum_{l=0}^{\infty}B_{l+1} n^l$ terminates after the second term, such that only the second virial coefficient $B_2$ needs to be calculated (the first virial coefficient takes the value $B_1 = 1$ and characterizes the classical ideal gas).
Let us consider a classical fluid of hard spheres in $d$ dimensions as the simplest case. An interaction parameter does not explicitly enter in this model, since the interaction is either zero if the hard spheres are not in contact, or infinite if they touch. The only coupling parameter is the radius $a$ of the hard spheres. The grand partition function $\mathcal{Z}=\exp (\beta PV)$
is expanded in powers of $z/\lambda^d$, where $z=\exp (\beta \mu)$ is the fugacity
and $\lambda=\hbar \sqrt{2\pi \beta/m}$ is the thermal wave length.
Each term of this ``Mayer cluster expansion'' can be expressed in terms of graphs \cite{Huang}.
Due to the linked cluster theorem $\mathcal{Z}$  reduces to a sum of connected graphs, $\mathcal{Z}_c$, with $\mathcal{Z}=\exp(\mathcal{Z}{_c}{-}1)$, which can be written as
\begin{equation}
\mathcal{Z}_c=\sum_{l=0}^{\infty}b_l \bigg( \frac{z}{\lambda^d} \bigg )^{\!l}.
\label{Z_c}
\end{equation}
Here the coefficients $b_l$ are determined by all possible connected graphs multiplied by a weight which is determined by configurational integrals over the volume. The typical scale of volume is the volume of a hard sphere with radius $a$ in $d$ dimensions, $v{_d}(a)=v{_d}(1) a^d $, where $v{_d}(1)=\pi{^{d/2}}/\Gamma(1{+}d/2)$ is the volume of a $d$-dimensional unit sphere.

\subsubsection{Equation of state}

By differentiating the grand potential $\Omega(T,V,\mu)=-\beta{^{-1}}\ln \mathcal{Z}=-PV$ with respect to the volume $V$\! or the chemical potential $\mu$, one obtains the pressure $P$ or particle number $N$, respectively, from which the equation of state is obtained by eliminating $z/\lambda^d$. In the limit $d \to \infty$ the evaluation of the coefficients $b_l$ simplifies significantly because for large $d$ every loop in a connected graph is suppressed exponentially by a factor $(\sqrt{3}/2){^d}/\sqrt{d}$. This is due to the fact that $v{_d}(1)$ decreases exponentially for $d \to \infty$, which implies that the cross section of the spheres vanishes. This leaves only loop-less (``tree'') diagrams.
As a consequence, only the second virial coefficient, given by $B_2=(1/2)v{_d}(a)$ remains, while all higher virial coefficients vanish in the limit $d \to \infty$.\footnote{This result for $B_2$ is plausible since $B_2$ has the dimension of a volume and $v{_d}(a)$ is the only characteristic volume in the problem.}
However, $B_2$ has also a $d$ dependence and vanishes for $d \to \infty$ since the volume of a $d$-dimensional unit sphere $v{_d}(1)$ goes to zero in this limit. Therefore a  scaling of $B_2$ with the dimension $d$ is required to reach a proper mean-field limit $d \to \infty$.\footnote{To this end a scaled (dimensionless) ``density per dimension'' $\rho$ was defined through $n v{_d}(a)=\rho^d$ in ref.~\cite{Frisch99}. However, thereby thermodynamic variables are scaled rather than the coupling parameter $a$ of the system, in contrast to the general construction principle of mean-field theories.} For this purpose we scale the hard-sphere radius $a$, the only coupling parameter in the problem, as $a=d^{\nu} a^{*}$, with $a^*$ as the scaled radius. The exponent $\nu$ has to be determined such that $B_2$ remains finite for $d \to \infty$, i.e., $v{_d}(a)= v{_d}( d^{\nu} a^*) \equiv v(a^*)=  {\rm const}$. With $\Gamma(1{+}d/2) \sim (d/2)^{d/2}$ for $d \to \infty$ one finds $\nu=1/2$, which implies the scaling
\begin{equation}
a \to \sqrt{d} a^* , a^* = {\rm const}
\label{Hardspherefluid-scaling}
\end{equation}
for $d \to \infty$. The shrinking of the volume (and thus of the cross section) of a $d$-dimensional unit sphere for increasing  $d$ is thus compensated by a corresponding increase of the radius of the hard sphere. Thereby $B_2$ becomes $B_2=(1/2)\, v(a^*)$. The mean-field equation of state of a classical hard sphere fluid then takes the form
\begin{equation}
\frac{PV}{k_B T}= N \left ( 1+ \frac12\, v(a^*)\, n  \right).
\label{eq-of-state}
\end{equation}
The interactions in the hard-sphere fluid are here described entirely by the second virial coefficient, which characterizes the interaction between two particles.

\section{Correlated electrons in solids\index{electronic correlations}}
\label{correlations}

\subsection{From the Ising model to the Hubbard model}\index{Ising model}\index{Hubbard model}

One of the most striking solid-state phenomena
is \emph{ferromagnetism}\index{ferromagnetism} as observed in magnetite (Fe$_3$O$_4$) and elemental iron (Fe).
How can ferromagnetism by explained?
A crucial first step in the development of a microscopic theory of ferromagnetism
was the formulation of the Ising model \cite{Ising} discussed in Sect.~\ref{sec:ising}.
Ising solved the model in $d=1$, found that a transition to a ferromagnetic phase does not occur, and concluded (incorrectly) that this holds also in $d=3$. The Ising model is a classical spin model. But it had already been shown by Bohr (1911) and van Leeuwen (1919) that magnetism is a quantum effect. Therefore another important step in the development of a theory of ferromagnetism was Heisenberg's formulation of a quantum spin model in 1928 \cite{Heisenberg}. With this model is was possible to explain the origin of the Weiss molecular field as the result of a quantum-mechanical exchange process.
Clearly, a model of localized spins cannot explain ferromagnetism observed in 3$d$ transition \emph{metals} such as iron, cobalt and nickel. As pointed out by Bloch \cite{Bloch} in 1929 an appropriate model had to include the mobile nature of the electrons, i.e., their wave character, which, in a solid, leads to electronic bands. The conditions for ferromagnetism which he obtained for free electrons where quite unrealistic.
%\footnote{For a historical review of the development of the quantum-mechanical theory of metals from 1928 to 1933, which describes the conceptual problems of that time, see ref.\cite{Hoddeson}.}
Obviously one had to go beyond free electrons and also take the mutual interaction between their electric charges into account. This immediately leads to an enormously difficult many-body problem, which is made particularly complicated by the fermionic nature of the electrons, their long-range Coulomb interaction, their high density in the metallic state, and the presence of a periodic lattice potential. Attempts by Slater \cite{Slater36} in 1936 to explain ferromagnetism in Ni by including the Coulomb interaction within Hartree-Fock theory were also not successful.

\vspace{-1ex}
\subsubsection{Electronic correlations}
\vspace{-1ex}

It became clear that one had to include \emph{correlations}, i.e., effects of the interaction between electrons which cannot be explained within a single-particle picture and therefore go beyond the physics described  by factorization approximations such as Hartree or Hartree-Fock mean-field theory. Wigner \cite{Wigner} was apparently the first who tried to calculate the contribution of the mutual electronic interaction to the ground state energy beyond the Hartree-Fock result, which he referred to as ``correlation energy''.\footnote{Electronic correlations play a fundamental role in modern condensed matter physics.
They are known to be strong in materials with partially filled $d$ and $f$ electron
shells and narrow energy bands, as in the 3$d$ transition metals or the rare--earths and their compounds,
%\footnote{A quasiclassical relation between the width of the energy band at the Fermi level and the degree of correlation can be derived as follows. Assuming that the correlated electrons  (or rather the quasiparticles, i.e., excitations) can be described by single-particle states ($\bm{k},\sigma$) with a dispersion $\epsilon_{\bm{k}}$, their velocity follows as $v_{\bm{k}} = |\nabla _{\bm{k}} \epsilon_{\bm{k}}|/\hbar$. Its value is roughly given by $v_{\bm{k}} \sim a/\tau$, where $a$ is  the lattice spacing and $\tau$ is the
%average time spent by an electron on an atom. Furthermore, the gradient of the energy can be estimated as $|\nabla_{\bm{k}} \epsilon_{\bm{k}}|\sim a W$, since $|\nabla_{\bm{k}}|\sim 1/k_F \sim a$, where $k_F$ is the Fermi wave number, and $|\epsilon_{\bm{k}}|$ is proportional to the band overlap $t$ and
%hence to the band width $W$. This then leads to the relation $\tau \sim \hbar/W$, which implies that the narrower the energy band, the longer an electron experiences the presence of other electrons, i.e.,  the stronger the correlations between the electrons.}
%
but they also occur in materials without rare-earth or actinide elements, such as ``Moir\'e'' heterostructures of van der Waals materials, e.g., twisted bilayers of graphene \cite{Moire} and
two-layer stacks of TaS$_2$
\cite{Vano}.
Electronic correlations  in solids lead to the emergence of complex behavior, resulting in rich phase diagrams.
In particular, the cooperation between the various degrees of freedom of the correlated electrons  (spin, charge, orbital momentum)
on a lattice with specific dimensionality and topology result in a wealth of correlation and ordering phenomena such as heavy fermion behavior \cite{Steglich}, high temperature superconductivity \cite{Schrieffer}, colossal magnetoresistance \cite{Dagotto}, Mott metal-insulator transitions\index{metal-insulator transition} \cite{tokura}, and Fermi liquid instabilities \cite{HvL}. The surprising discovery of a multitude of correlation phenomena in  Moir\'e heterostructures \cite{Moire} reinforced this interest in correlation physics.
The unusual properties of correlated electron systems are not only of interest for fundamental research but may also be relevant for technological applications.
Indeed, the exceptional sensitivity of correlated electron materials upon changes of external parameters such as temperature, pressure, electromagnetic fields, and doping can be employed to develop materials with useful functionalities \cite{Functionality}. In particular, Moir\'e heterostructures\index{Moir\'e heterostructures} may enable ``twistronics'',\index{twistronics} a new approach to device engineering \cite{Twistronics}.
Consequently there is a great need for the development of appropriate models and theoretical investigation techniques which allow for a comprehensive, and at the same time reliable, exploration of correlated electron materials  \cite{Juelich-lecture-notes}.}
Such an approach is faced by two intimately connected problems: the need for a sufficiently simple model of correlated electrons which unifies the competing approaches by Heisenberg and Bloch (namely the picture of localized and itinerant electrons, respectively),
%the Heitler-London method for molecules and the molecular orbital method,
and its solution within a more or less controlled approximation scheme. Progress in this direction
was slow.\footnote{One reason for the slow development was that in the nineteen-thirties and forties nuclear physics attracted more attention than solid-state physics, with a very specific focus of research during the 2nd World War. But apart from that, the sheer complexity of the many-body problem itself did not allow for quick successes. High hurdles had to be overcome, both regarding the development of appropriate mathematical techniques (field-theoretic and diagrammatic methods, Green functions, etc.) and physical concepts (multiple scattering, screening of the long-range Coulomb interaction, quasiparticles and Fermi liquid theory, electron-phonon coupling, superconductivity, metal-insulator transitions,\index{metal-insulator transition} disorder,\index{disorder} superexchange, localized magnetic states in metals, etc.). A discussion of the many-body problem and of some of the important developments up to 1961 can be found in the lecture notes and reprint volume by Pines \cite{Pines-book}.\label{Many-body}}
A microscopic model of ferromagnetism in metals (``band ferromagnetism'')
did not emerge
until 1963, when a model of correlated lattice electrons was proposed independently by Gutzwiller \cite{Gutzwiller}, Hubbard \cite{HubbardI}, and Kanamori \cite{Kanamori} to explain ferromagnetism in 3$d$ transition metals.
Today this model is referred to as ``Hubbard model''  \cite{Montorsi}.

\begin{figure}
\centerline{\includegraphics[width=0.55\textwidth]{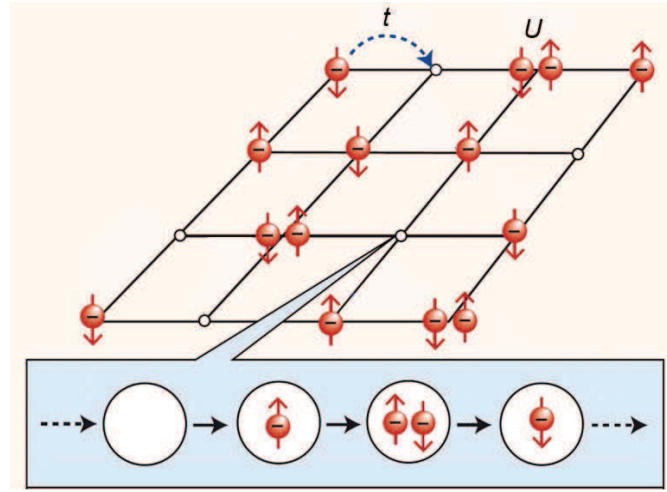}}
\caption{Schematic illustration of interacting electrons in a solid described by the Hubbard model.\index{Hubbard model}  The ions enter only as a rigid lattice, here represented by a square lattice ($d=2$, $Z=4$). The electrons, which have mass, negative charge, and spin ($\uparrow$   or $\downarrow$), are quantum particles which ``hop'' from one lattice site to the next with a hopping amplitude $t$. Together with the lattice structure this determines the band structure of the non-interacting electrons. The quantum dynamics leads to fluctuations in the occupation of lattice sites as indicated by the sequence: a lattice site can either be unoccupied, singly occupied ($\uparrow$    or   $\downarrow$), or doubly occupied.  When two electrons meet on a lattice site, which is only possible if they have opposite spins because of the Pauli exclusion principle, they encounter a local interaction $U$.}
\label{Hubbard_model}
\end{figure}

The single-band Hubbard model\index{Hubbard model} is the minimal microscopic lattice model of interacting fermions with local interaction.\footnote{It is interesting to note that Anderson had introduced the main ingredient of the Hubbard model, namely a local interaction between spin-up and spin-down $d$ electrons with strength $U$\!, already in his 1959 paper on the theory of superexchange interactions \cite{Anderson59} and, even more explicitly in his 1961 paper on localized magnetic states in metals, where he formulated a model of $s$ and $d$ electrons referred to today as ``single impurity Anderson Model'' (SIAM)\index{single-impurity Anderson model} or ``Anderson impurity model'' (AIM) \cite{Anderson61}. The latter paper inspired Wolff \cite{Wolff61} to study the occurrence of localized magnetic moments in dilute alloys with a single-band model of noninteracting $d$ electrons which interact on a single site. In this sense the Hubbard model could be called ``periodic Wolff model''\index{Wolff model} in analogy to the standard terminology ``periodic Anderson model'', which generalizes the AIM by extending the interaction to all sites of the lattice. Apparently Gutzwiller, Hubbard and Kanamori did not know (or
were not influenced) by the earlier work of Anderson and Wolff; at least they did not reference these papers.}
The Hamiltonian consists of
the kinetic energy $\hat{H}_0$ and
the interaction energy $\hat{H}_{\rm int}$ (in the following operators are
denoted by a hat):
\begin{subequations}
\label{G11.7}
\begin{eqnarray}
\hat{H} & = & \hat{H}_0 + \hat{H}_{\rm int}, \\[10pt]
\label{G11.7a}
\hat{H}_0 & = & \sum_{\bm{R}_i , \bm{R}_j} \sum_{\sigma}
t_{ij}  \hat{c}_{i \sigma}^\dagger \hat{c}_{j \sigma}^{} = \sum_{\bm{k} , \sigma}
\epsilon_{\bm{k}} \hat{n}_{\bm{k} \sigma}^{}, \label{G11.7b} \\[10pt]
\hat{H}_{\rm int} & = & U \sum_{\bm{R}_i} \hat{n}_{i \uparrow} \hat{n}_{i \downarrow}  \equiv  U \hat{D}.
\label{G11.7c}
\end{eqnarray}
\end{subequations}
Here $\hat{c}_{i \sigma}^\dagger (\hat{c}_{i \sigma}^{})$ are creation
(annihilation) operators of fermions with spin $\sigma$ at site  $\bm{R}_i$
(for simplicity denoted by $i$), $\hat{n}_{i \sigma}^{} = \hat{c}_{i
\sigma}^\dagger  \hat{c}_{i \sigma}^{}$, and $\hat{D}=\sum_{\bm{R}_i}\hat{D_i}$  is the operator of total double occupation of the system with $\hat{D_i}=\hat{n}_{i \uparrow} \hat{n}_{i \downarrow}$ as the operator of double occupation of a lattice site $i$. The Fourier transform of the kinetic
energy  in \eqref{G11.7b}, where $t_{ij}$ is the amplitude for hopping
between sites $i$ and $j$, defines  the dispersion $\epsilon_{\bm{k}}$ and
the momentum distribution  operator $\hat{n}_{\bm{k} \sigma}^{}$.
In the following the
hopping is restricted to nearest-neighbor sites
 $i$ and $j$, such that
$ - t \equiv t_{ij}$.
A schematic picture of the Hubbard model is shown in Fig.~\ref{Hubbard_model}.

\subsection{Characteristic features of the Hubbard model}

In the Hubbard model the Coulomb interaction between two electrons is assumed to be so strongly
screened that it can be described by a purely local interaction which occurs only \emph{on} a lattice site.\footnote{Thereby the Hubbard model applies particularly well
%to lattice fermions with a point interaction, such as
to cold fermionic atoms
in optical lattices\index{cold atoms in optical lattices} where the bare interaction is indeed extremely short-ranged \cite{cold-atoms2}.}
Due to the Pauli principle a local interaction is only possible if the two electrons have opposite spin.\footnote{It appears as if the interaction between the electrons was spin-dependent. But the Coulomb interaction is, of course, a spin-independent two-body
interaction. The fact that the operators in \eqref{G11.7c} contain spin indices is merely a consequence of
the occupation-number formalism.} This distinguishes a local interaction from other model interactions since it has no classical counterpart.
The interaction is therefore completely independent of the lattice structure and spatial dimension of the system.
The  kinetic energy $\hat{H}_0$ is diagonal in momentum space and reflects the wave nature of the electrons, while the interaction energy $\hat{H}_{\rm int}$ is diagonal in position space and characterizes their particle nature.
In view of the uncertainty principle the two terms are therefore extremely ``quantum incompatible''.

The physics described by the Hubbard model is very different from that of electrons with a long-range Coulomb interaction in the continuum. Therefore the Hubbard model is far from obvious.
  Its formulation required fundamentally new insights into the nature of the many-body problem of interacting fermions (see footnote \ref{Many-body}). In particular, screening is a basic ingredient of the many-body problem of metals.

A direct interaction between electrons with equal spin direction, e.g., on neighboring sites, is not described by the model, but can be easily included. Similarly the model can be generalized to more than one band. Indeed, in Wannier basis the Hubbard model can be  derived systematically  from a general Hamiltonian of interacting electrons including the kinetic energy, ionic potential $U_{\text {ion}}(\bm{r})$, and two-body Coulomb interaction $V_{\text{ee}}(\bm{r}{-}\bm{r}^{\prime})$ \cite{HubbardI}.\footnote{In Wannier basis the general Hamiltonian
%, $\hat{H}
%  =
%  \sum_{ij}t_{ij}^{\alpha}
%  \hat{c}_{i\alpha\sigma}^{\dagger}
%  \hat{c}_{j\alpha\sigma}^{\vphantom{\dagger}}
%  +
%  \frac{1}{2}
%  \sum_{ijmn}v_{ijmn}^{\alpha\beta\mu\nu}
%  \hat{c}_{i\alpha\sigma^{\dagger}}
%  \hat{c}_{j\beta\sigma^{\prime}}^{\dagger}
%  \hat{c}_{n\nu\sigma^{\prime}}^{\vphantom{\dagger}}
%  \hat{c}_{m\mu\sigma}^{\vphantom{\dagger}}$,
has infinitely many bands and model parameters.
%; here, $\hat{c}_{i\alpha\sigma}$ annihilates an electron with spin $\sigma$ in a Wannier orbital $\alpha$
% localized at site $i$, etc.
In the one-band model \eqref{G11.7} all other bands are projected
onto a single effective $s$ band. The matrix elements of the kinetic energy
%, $t_{ij}^{\alpha}=\langle i\alpha|-\frac{1}{2m}\nabla^{2}+U_{\text {ion}}(\bm{r})|j\alpha\rangle$,
and the Coulomb interaction
%, $v_{ijmn}^{\alpha\beta\mu\nu}
%  =
%  \left\langle i\alpha,j\beta|V_{\text{ee}}(\bm{r}-\bm{r}^{\prime})|m\mu,n\nu\right\rangle$,
are expected to fall off quickly with distance. Therefore
one usually retains only the first few contributions. Thus
hopping is restricted to nearest-neighbor sites
 $i$ and $j$, such that
$ - t \equiv t_{ij}$.
It is also natural to assume that the local part of the interaction (``Hubbard $U$'')
%$U \equiv v_{iiii}$
is the
largest matrix element of the Coulomb interaction. Keeping
only these parameters
%$t$ and $U$
one obtains the Hubbard model, \eqref{G11.7a}, \eqref{G11.7c}.
However, nearest-neighbor interactions (density-density, bond-charge, exchange interactions and hopping of doubly occupied sites)
may be of appreciable
size \cite{HubbardI}.
%These are the two-site terms of the interaction
%$V \equiv v_{ijij}$ (density-density interaction), $X \equiv v_{iiij}$ (bond-charge interaction), $F \equiv v_{ijji}$ (exchange interaction), $F' \equiv v_{iijj}$ (hopping of doubly occupied sites), where $i$ and $j$ are
%nearest neighbor sites.
For a detailed discussion see ref.~\cite{Kollar96}. }

Since the single-band Hubbard model is obtained from a general interaction Hamiltonian it is the fundamental quantum lattice model of interacting fermions\footnote{More generally, the Hubbard model is the fundamental lattice model of quantum particles, since it may also be used for interacting bosons \cite{cold-atoms1,cold-atoms2}.}. As a consequence, many well-known models can be derived from the Hubbard model in certain limits of the model parameters. For example, at half filling and in the limit $U \gg t$ the Hubbard model corresponds to the Heisenberg model with antiferromagnetic exchange coupling $J=4t{^2}/U$\!.\,\footnote{This can be shown using second order degenerate perturbation theory or by a unitary (Schrieffer-Wolff) transformation.}

\subsubsection{Does the Hubbard model explain band ferromagnetism?}\index{ferromagnetism}

As discussed above, the Hubbard model  was originally introduced to provide a microscopic explanation of ferromagnetism in 3$d$ transition metals \cite{Gutzwiller,HubbardI,Kanamori}. Indeed, the Hubbard interaction favors a ferromagnetic ground state,
since this corresponds to the state with the lowest energy (zero energy) due to the absence of doubly occupied sites.
However, this argument ignores the kinetic energy. While the lattice structure and spatial dimension do not influence the Hubbard interaction, both play an important role in the  kinetic energy, where they determine, for example, the density of states of the electronic band at the Fermi energy (see section \ref{Band-ferromagnetism}).

In spite of the extreme simplifications of the Hubbard model compared  with interacting electrons in a real solid, the model still cannot be solved  analytically, except in dimension $d=1$ for nearest-neighbor hopping
\cite{Lieb+Wu}. For dimensions $d = 2,3$, approximations are required.\footnote{In view of the complexity of the many-body problem, progress in this field often relies on making good approximations. As Peierls wrote: ``\emph{... the art of choosing a suitable approximation, of checking its consistency and finding at least intuitive reasons for expecting the approximation to be satisfactory, is much more subtle than that of solving an equation exactly.}'' \cite{Peierls}.} Here mean-field theories play an important role.

\section{Static mean-field theories of the Hubbard model}\index{mean-field theory}\index{Hubbard model}

\subsection{Hartree approximation}
\label{sec:Hartree}

Lattice fermion models such as the Hubbard model are much more complicated than models with localized spins.
Therefore the construction of a mean-field theory  with the
comprehensive properties of the mean-field theory of the Ising
model will be more complicated, too.
The simplest static mean-field theory of the Hubbard
model is the Hartree approximation \cite{Langer+Dichtel}. To clarify the characteristic features of this mean-field theory we proceed as in the derivation of the mean-field theory  of the Ising model and factorize the interaction term. To this end we rewrite the Hubbard interaction in the form of \eqref{G11.1a}, i.e., we let an electron with spin $\sigma$ at site $\bm{R}_i$ interact with a local field $\hat{h}_{i \sigma}$ (an operator, which has a dynamics) produced by an electron with opposite spin on that site:

\begin{equation}
\hat{H}_{\rm int} =
\sum\limits_{\bm{R}_i}
\sum\limits_{\sigma}
\hat{h}_{i \sigma}\hat{n}_{i \sigma},
\label{G14.1}
\end{equation}
where $\hat{h}_{i \sigma}=\frac{1}{2} U \hat{n}_{i -\sigma}$ (the factor $1/2$ is due to the summation over both spin directions).
Next we
replace the operator $\hat{h}_{i \sigma}$ by its expectation value $\langle \hat{h}_{i \sigma} \rangle$, now a real number, and obtain the mean-field Hamiltonian
\begin{equation}
\hat{H}^{\rm MF} = \hat{H}_{\rm kin} + \sum_{\bm{R}_i , \sigma}
\langle \hat{h}_{i \sigma} \rangle \hat{n}_{i \sigma}^{}  + E_{\rm shift},
\label{G14.2}
\end{equation}
where $E_{\rm shift}$ is a constant energy shift.
Now a $\sigma$-electron at site
$\bm{R}_i$ interacts only with a local \emph{static} field
$\langle \hat{h}_{i \sigma} \rangle
= \frac{1}{2} U \langle \hat{n}_{i -\sigma} \rangle$.
The above decoupling of the operators
 corresponds to the Hartree approximation\footnote{Since the Hubbard interaction acts only between electrons with opposite spin on the same lattice site an exchange (Fock) term does not arise.}
whereby correlated fluctuations \emph{on} the site
$\bm{R}_i$ are neglected.

It should be noted that although \eqref{G14.2} is now an effective one-particle
problem it cannot be solved exactly since, in general,
 the  mean field
$\langle \hat{h}_{i \sigma} \rangle$ may vary from site to site, leading to solutions without long-range order.
This is a new feature originating from the quantum-mechanical kinetic energy in the Hamiltonian.

The Hartree approximation is valid in the weak-coupling ($U{\to}0$) and/or low-density limit ($n{\to}0$), but clearly does not become exact in the limit $d \to \infty$, since
the Hubbard interaction between two electrons is purely local and hence does not dependent on the spatial dimension. Therefore the physics behind the factorizations \eqref{G11.4} and \eqref{G14.2} is very different. Namely, \eqref{G11.4} describes the decoupling of a spin from a bath of infinitely many neighboring spins whose fluctuations become unimportant in the limit $d \to \infty$, while \eqref{G14.2} corresponds to the decoupling of an electron from  \emph{one} other electron (with opposite spin) on the same site.

While the Hartree approximation is useful for investigations at weak coupling, it will lead to fundamentally wrong results at strong coupling, when the
double occupation of a lattice site becomes energetically highly unfavorable and is therefore suppressed. Indeed, a factorization of the local correlation function $\langle \hat{n}_{i \uparrow}\, \hat{n}_{i \downarrow}  \rangle \to \langle \hat{n}_{i\uparrow} \rangle \langle
\hat{n}_{i\downarrow}  \rangle $ eliminates correlation effects generated by the local quantum dynamics.
Hence the nature of the Hartree mean-field theory  of \mbox{spin-$\frac{1}{2}$} electrons with a local interaction is very different from the mean-field theory  of spins with nearest-neighbor coupling.

\subsection{Gutzwiller approximation}\index{Gutzwiller wave function}\index{Gutzwiller approximation}
\label{Gutzwiller-approximation}

 Another useful approximation scheme for quantum many-body systems
   makes use of variational wave functions \cite{VWF}. Starting from a physically motivated many-body trial wave function the energy expectation value is calculated and then minimized with respect to the variational parameters. Although variational wave functions usually yield only approximate results, they have several advantages: they are physically intuitive, can be custom tailored to a particular problem, can be used even when standard perturbation methods fail or are inapplicable, and provide a rigorous upper bound for the exact ground state energy by the Ritz variational principle.

 To investigate the properties of the electronic correlation model which he had introduced \cite{Gutzwiller} (but which was later named after Hubbard), Gutzwiller also  proposed a simple variational wave function \cite{Gutzwiller}. This ``Gutzwiller wave function'' introduces correlations into the wave function for  non-interacting particles by a purely local correlation factor in real space, which is constructed from the double occupation operator $\hat{D}$, \eqref{G11.7c}, as
  \begin{subequations}
\label{GWF}
\begin{eqnarray}
|\Psi_G \rangle & = & g^{\hat{D}} \, |{\rm FS} \rangle \\[10pt]
 & = & \prod_{\bm{R}_i} \big( 1 - (1{-}g) \hat{D_i} \big)\,  |{\rm FS} \rangle.
\end{eqnarray}
  \end{subequations}
  Here $|{\rm FS} \rangle$ is the wave function of the non-interacting Fermi sea
   %(\underline{F}ermi \underline{G}as),
   and $g$ is a variational parameter with $0 \leq g \leq 1$.
  The projector  $g^{\hat{D}}$ globally reduces the amplitude of those spin configurations
in $|{\rm FS} \rangle$ with too many doubly occupied sites for given repulsion $U$\!.\; The limit
$g = 1$ describes the non-interacting case, while
$g \rightarrow 0$
corresponds to $U \rightarrow \infty$.
The Gutzwiller wave function can be used to calculate the expectation value of an operator,
e.g., the ground state energy
of the Hubbard model, \eqref{G11.7}, as
\be
E_{\rm G}(g,U) \equiv
\frac{\langle \Psi_{G} | {\Hop} | \Psi_{G} \rangle }
{\langle \Psi_{G} | \Psi_{G} \rangle}.
\ee
By computing the minimum of $E_{\rm G}(g,U)$ with respect to the variational parameter  $g$, the latter is determined as a function of the interaction parameter $U$\!.

In general the evaluation of expectation values in terms of $| \Psi_G \rangle$ cannot be performed exactly. Therefore Gutzwiller introduced a non-perturbative approximation scheme whereby he obtained an explicit expression for the ground state energy of the Hubbard model \cite{Gutzwiller,Gutzwiller1965,Fazekas}.
The results of Gutzwiller's rather complicated approach were later re-derived by
 counting classical spin configurations \cite{Ogawa}; for details see ref.~\cite{Vollhardt84} and section 3 of ref.~\cite{Julich-2014}. The Gutzwiller approximation is therefore a semiclassical approximation.
In the next section we will see that it can also be derived by
calculating the expectation values of operators in terms of the Gutzwiller wave function in the limit $d \to \infty$.

\subsubsection{Brinkman-Rice metal-insulator transition }\index{metal-insulator transition}
\label{Brinkman-Rice}

The results of the Gutzwiller approximation \cite{Gutzwiller,Gutzwiller1965} describe a correlated, normal-state fermionic system
at zero temperature whose momentum distribution has a discontinuity $q$ at the Fermi level, with $q=1$ in the non-interacting case, which is reduced to $q<1$ by the interaction as in a Landau Fermi liquid.\index{Fermi liquid} In 1970 Brinkman and Rice \cite{BR70} noticed that in the case of a half-filled band ($n_{\uparrow}=n_{\downarrow}= 1/2$) the Gutzwiller approximation describes a transition at a finite critical interaction strength $U_c$ from an itinerant to a localized state, where lattice sites are singly occupied and the discontinuity $q$ vanishes. This ``Brinkman-Rice transition'' therefore corresponds to a correlation induced (``Mott'') metal-insulator transition.\index{Mott transition} They argued  \cite{BR70} that the inverse of $q$ can be identified with the effective mass of Landau quasiparticles, $q^{-1}=m^*/m \geq 1$,  which  diverges at $U_c$.

The Gutzwiller approximation yields results which are physically very reasonable. In the 1970s and 80s it was the only approximation scheme which was able to describe a Mott metal-insulator transition at a finite value of the interaction \emph{and} was in accord with basic properties of Landau Fermi liquid theory.\footnote{Other well-known approximation schemes, in particular those proposed by Hubbard, do not have these important properties: in the Hubbard-I approximation \cite{HubbardI}, which interpolates between the atomic limit and the non-interacting band, a band gap opens for any  $U>0$, while the  Hubbard-III approximation \cite{HubbardIII}, which corresponds to the coherent potential approximation  \cite{CPA}\index{coherent potential approximation (CPA)} for disordered systems,\index{metal-insulator transition} the Fermi surface volume is not conserved. }
This was confirmed by a detailed investigation of the assumptions and implications of the Gutzwiller approximation, where I showed
that the Gutzwiller-Brinkman-Rice theory was not only in qualitative \cite{AB78}, but even in good quantitative agreement with experimentally measured properties of normal-liquid $^{3}$He \cite{Vollhardt84}; for a discussion see section 3 of~ref.~\cite{Julich-2014}.

\subsubsection{Can the Gutzwiller approximation be derived by quantum many-body methods?}

The semiclassical Gutzwiller approximation clearly has features of a mean-field theory, since the kinetic energy of the correlated system is merely a renormalization of  the kinetic energy of the non-interacting system.
This also explains why the results obtained for the Hubbard lattice model can even be used to describe liquid $^{3}$He \cite{Vollhardt84,VWA87}. However, the validity of the Gutzwiller approximation was unclear for a long time.
In particular, it was not known how to improve the approximation systematically.
This was clarified a few years later, when the Gutzwiller approximation was re-derived in two different ways: as a slave-boson mean-field theory \cite{KR86}, and by an exact diagrammatic calculation of expectation values of operators in terms of the Gutzwiller wave function in the limit $d\to\infty$ \cite{MV87+MV88,MV89,metzner89a}\footnote{Kotliar and Ruckenstein  \cite{KR86}
formulated a functional integral representation of the Hubbard and Anderson models in terms of auxiliary bosons, whose simplest saddle-point approximation (``slave-boson mean-field theory'') reproduces the results of the Gutzwiller approximation. Thus they showed that the results of the Gutzwiller approximation\index{Gutzwiller wave function} can also be obtained without the use of the Gutzwiller variational wave function.\index{Gutzwiller wave function} By applying a gauge transformation to the Gutzwiller wave function Gebhard \cite{Gebhard90} later found that the calculation of expectation values with this wave function can be performed in $d = \infty$ even without diagrams. This provided a direct link between the slave-boson mean-field theory and the results obtained with the Gutzwiller wave function in $d = \infty$. The latter approach was generalized by Gebhard and collaborators to multi-band Hubbard models, which can be used to describe the effect of correlations in real materials  (``Gutzwiller density functional theory'') \cite{GDFT1+GDFT3}.}. The latter derivation will be discussed~next.

\section{Lattice fermions in infinite dimensions\index{infinite dimensions}}

The expectation values of the kinetic and the interaction energy of the Hubbard model \eqref{G11.7}  can, in principle, be calculated in terms of the Gutzwiller wave function\index{Gutzwiller wave function} for arbitrary dimensions $d$, using diagrammatic quantum many-body theory. Introducing a new analytic approach in which expectation values are expressed by sums over different lattice sites such that Wick's theorem leads to contractions which involve only anticommuting numbers, Walter Metzner and I showed that in $d=1$
the diagrams can be calculated analytically to all orders \cite{MV87+MV88};\footnote{The diagrams have the same form as the usual Feynman diagrams in quantum many-body theory,
% (due to the locality of the interaction they are identical to those of a $\Phi^4$ theory)
but a line corresponds to the time-independent one-particle density matrix $ g_{ij , \sigma}^0 = \langle \hat{c}_{i \sigma}^\dagger \hat{c}_{j \sigma}^{} \rangle_0$ of the non-interacting system rather than to the one-particle propagator $G_{ij , \sigma}^0 (t)$ since the variational approach involves only static quantities.}\label{one-particle-density-matrix}
for a discussion see section 4 of ref.~\cite{Julich-2014}.
We also
observed  that the values of these diagrams could be calculated in $d =\infty$ when the momentum conservation at a vertex was neglected.
When summing over all diagrams this approximation gave exactly the results of the Gutzwiller approximation \cite{MV87+MV88}. Thus the Gutzwiller approximation\index{Gutzwiller approximation} had been derived systematically by diagrammatic quantum many-body methods
in $d=\infty$.
This showed that the limit $d\rightarrow \infty$ was not only useful for the investigation of spin models, but also in the case of lattice fermion models.\footnote{In the limit $d\rightarrow \infty$ simplifications occur even for fermions in the continuum. In the ground state of a $d$-dimensional Fermi gas in the continuum, spatial isotropy implies that $\bm{k}$-states occupy a Fermi sphere. As is well-known from the discussion of the classical ideal gas in the microcanonical ensemble in the limit $d \to \infty$ the volume of a high-dimensional sphere is located essentially \emph{at} the surface. Therefore in the high dimensional Fermi sphere essentially all states lie at the Fermi energy $E_F$.
This is easily confirmed by calculating the energy per particle in $d$ dimensions, ${E_0}^{(d)}/N = \big(d/(d{+}2)\big) E_F$ \cite{Acharyya}, which indeed gives ${E_0}^{(\infty)}/N=E_F$ in  $d = \infty$. It implies that the pressure of the Fermi gas at $T=0$ (``Fermi pressure''), $P=2{E_0}^{(d)}/Vd$, which is larger than zero in finite dimensions, goes to zero in this limit.\label{d-infinity-continuum}}

\subsection{Simplifications of diagrammatic quantum many-body theory}
%\label{sec:scaling}

The simplifications of diagrammatic many-body perturbation theory arising in the limit $d \to \infty$
are due to a collapse of irreducible diagrams in position space, which implies that only site-diagonal (``local'') diagrams, i.e., diagrams which only depend on a single site, remain   \cite{MV89,metzner89a}.
%
%In particular, the one-particle irreducible self-energy\index{self-energy} is then also completely local.
%
To understand the reason for this diagrammatic collapse let us, for simplicity, consider diagrams where lines correspond to the one-particle density matrix $ g_{ij , \sigma}^0$ as they enter in the calculation of expectation values with the Gutzwiller wave function (note that, since $g_{ij,\sigma}^0 =\lim_{t\rightarrow 0^-} G^0_{ij,\sigma}(t)$, the following arguments are equally valid for the one-particle Green function $G_{ij, \sigma}^0 (t)$ or its Fourier transform).

The one-particle density matrix $g_{ij,\sigma}^0$ may be interpreted as the quantum amplitude of the hopping of an electron with spin $\sigma$
between sites $\bm{R}_i$ and $\bm{R}_j$. The square of its magnitude is therefore proportional to the \emph{probability} of an electron to hop from $\bm{R}_j$ to a site $\bm{R}_i$. For nearest-neighbor sites $\bm{R}_i$, $\bm{R}_j$ on a lattice with coordination number $Z$ this implies $| g_{ij , \sigma}^0 |^2 \sim {\cal O}(1/Z)$, such that on a hypercubic lattice, where $Z=2d$, and large $d$ one finds  \cite{MV89,metzner89a}
\begin{equation}
g_{ij, \sigma}^0 \sim {\cal O} \Big( \frac{1}{\sqrt{d}} \Big),
\label{G11.20}
\end{equation}
and for general $i, j$ one obtains
\cite{metzner89a,vanDongen89}
\begin{equation}
g_{ij , \sigma}^0 \sim {\cal O}
\Big( 1/d^{\| \bm{R}_i - \bm{R}_j \|/2 } \Big).
\label{G11.22}
\end{equation}
Here $\| \bm{R} \| = \sum_{n = 1}^{d} | R_n | $ is the length of $\bm{R}$ in
the ``Manhattan metric'', where electrons
only hop along horizontal or vertical lines, but never along a diagonal; for further discussions of diagrammatic simplifications see ref.~\cite{Vollhardt-Salerno}.

For non-interacting electrons at $T = 0 $ the expectation value of the kinetic energy is given by
\begin{equation}
E_{\rm kin}^0 = - t \sum_{\langle \bm{R}_i, \bm{R}_j \rangle} \sum_{\sigma}
 \; g_{ij, \sigma}^{0}.
\label{G11.19}
\end{equation}
The sum over nearest neighbors leads to a factor  ${\cal O} (Z)$ (which is ${\cal O} (d)$ for a hypercubic lattice). In view of the $1/\sqrt{d}$ dependence of $g_{ij, \sigma}^0$ it is therefore necessary to scale the nearest-neighbor hopping amplitude $t$ as \cite{MV89,metzner89a}
\begin{equation}
t = \frac{t^*}{\sqrt{d}}, \; t^* = {\rm const.},
\label{G11.11}
\end{equation}
so that the kinetic energy remains finite for $d\rightarrow \infty$.
The same result may be derived in a momentum-space formulation.\footnote{The need for the scaling \eqref{G11.11} also follows from the density of states of non-interacting electrons. For nearest-neighbor hopping on a $d$-dimensional hypercubic lattice, $\epsilon_{\bm{k}}$ has the form
$\epsilon_{\bm{k}} = - 2t \sum_{i = 1}^{d} \; \cos k_i$ (here and in the following we set Planck's constant $\hbar $, Boltzmann's constant $k_{B}$, and the lattice spacing equal to unity). The density of states corresponding to $\epsilon_{\bm{k}}$ is given by
$N_d (\omega) = \sum_{\bm{k}} \delta (\omega{-}\epsilon_{\bm{k}})$, which is the probability density for finding the value $\omega = \epsilon_{\bm{k}}$
for a random choice of $\bm{k} = (k_1, \ldots, k_d)$. If the momenta $k_i$ are
chosen randomly, $\epsilon_{\bm{k}}$  is the sum of $d$ many
independent (random) numbers $-2t \cos k_i$. The central limit theorem then implies that in the limit $d \to \infty$ the density of states is given by a Gaussian, i.e.,  $N_d (\omega) \stackrel{d \to \infty}{\longrightarrow} \; \frac{1}{2t \sqrt{\pi d}}
 \exp\!\Big(\!{-}\big( \frac{\omega}{2t \sqrt{d}} \big)^2 \Big)$.
Only if $t$ is scaled with $d$  as  in \eqref{G11.11}
 does one obtain a non-trivial density of states $N_{\infty} (\omega)$ in $d= \infty$ \cite{Wolff83,MV89} and thus a finite kinetic energy. The density of states on other types of lattices in $d= \infty$ can be calculated similarly \cite{General_DOS,FM-Ulmke}.}
It is important to bear in mind that, although $g_{ij , \sigma}^0 \sim 1/\sqrt{d}$ vanishes
for $d \to \infty$, the electrons are still mobile.
Indeed,
 even in the limit
$d \to \infty$ the off-diagonal elements of $g_{ij , \sigma}^0$ contribute, since electrons may
hop to $Z \sim {\cal O} (d)$ many nearest neighbors with amplitude $t^*/\sqrt{d}$.
\footnote{\label{GA-mean-field}
Due to the diagrammatic collapse in $d=\infty$ it is possible to sum the diagrams representing the expectation value of operators in terms of the Gutzwiller wave function\index{Gutzwiller wave function} exactly \cite{MV89,metzner89a}. Lines in a diagram correspond to the one-particle density matrix $ g_{ij , \sigma}^0$ rather than the one-particle Green's function $G^0_{ij,\sigma}(t)$. The results obtained in infinite dimensions are found to coincide with those of the Gutzwiller approximation,\index{Gutzwiller wave function} which was originally derived by classical counting of spin configurations (section \ref{Gutzwiller-approximation}). Obviously the Gutzwiller approximation is a mean-field theory\index{mean-field theory} -- but what is the corresponding mean field?\index{mean field} To answer this question we note that the one-particle irreducible self-energy\index{self-energy} (i.e., the quantity which has the same topological structure as the self-energy in the Green's function formalism) and which is denoted by $S^*$ in \cite{MV89,metzner89a}, becomes site-diagonal in infinite dimensions. In the translationally invariant, paramagnetic phase the self-energy is then site-independent, i.e., constant. Since the self-energy encodes the effect of the interaction between the particles (see footnote \ref{Selfenergy-Gutz}) it may be viewed as the \emph{global mean field} of the Gutzwiller approximation; its value is given by eq.~(10) in \cite{MV89}. Expectation values can be evaluated exactly even when the Fermi sea $|{\rm FS} \rangle$ in \eqref{GWF} is replaced by an arbitrary, not necessarily translationally invariant one-particle starting wave function $| \Phi_0 \rangle$ \cite{MV89,metzner89a}. In general, the mean field $S_{ii}^{*}$ then depends on the lattice site $i$ (but is still site-diagonal) and is determined by the self-consistent eq.~(11)\index{self-consistent equation} in \cite{MV89}, which reduces to eq.~(10) in the paramagnetic case. Details of the calculation of $S_{ii}^{*}$ can be found in the paper by Metzner \cite{metzner89a}.}

\subsection{The Hubbard model in \texorpdfstring{$d\,{=}\,\infty$}{d=infinity}}\index{Hubbard model}
\label{sec:scaling}

\begin{figure}
\centerline{\includegraphics[width=0.55\textwidth]{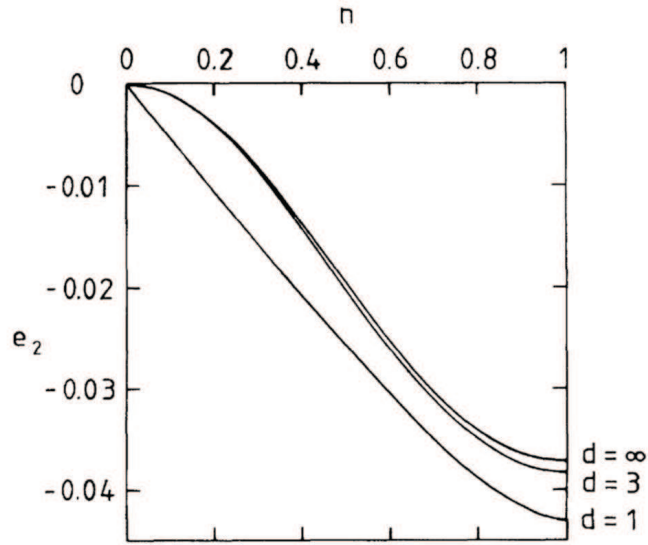}}
\caption{Correlation energy  of the Hubbard model in second-order Goldstone perturbation theory in $U$ in units of $2U^2/|\epsilon_0|$ vs.\ density $n$ in dimensions $d = 1,3, \infty$. Here $\epsilon_0 $
is the kinetic energy for $U=0$ and $n = 1$; adapted from ref.~\cite{MV89}.}
\label{weak-coupling-correlation-energy}
\end{figure}

A rescaling of the microscopic parameters of the Hubbard model with $d$ is only required in the kinetic energy, since the interaction term is independent of the spatial dimension.\footnote{In the limit $d\to\infty$, interactions beyond the Hubbard interaction,
e.g., nearest-neighbor interactions such as
$\hat{H}_{nn} = \sum_{\langle \bm{R}_i , \bm{R}_j \rangle} \; \sum_{\sigma \sigma'} \; V_{\sigma \sigma'} \hat{n}_{i \sigma}^{}
\hat{n}_{j \sigma'}^{}$
\label{G11.30}
have to be scaled, too. %, in the limit $d \to \infty$.
In this case a  scaling  as in the Ising model,
$V_{\sigma \sigma'} \to V_{\sigma \sigma'}^*/Z$,
is required \cite{MH89a}.
In $d = \infty$ non-local contributions therefore reduce to their (static) Hartree substitute
and only the
Hubbard interaction remains dynamical.} Altogether this implies that only the Hubbard Hamiltonian with a rescaled kinetic energy
\begin{equation}
\hat{H} = - \frac{t^*}{\sqrt{d}} \; \sum_{\langle \bm{R}_i, \bm{R}_j \rangle}
\sum\limits_\sigma
 \; \hat{c}_{i \sigma}^\dagger \hat{c}^{}_{j\sigma}
+ U \sum_{\bm{R}_i} \hat{n}_{i \uparrow} \hat{n}_{i \downarrow}
\label{scaled-HM}
\end{equation}
 has a non-trivial
 $d \to \infty$ limit where
 both  the kinetic energy and the interaction contribute.
 Namely, it is the \emph{competition} between the two terms which leads to interesting many-body physics.
Most importantly, the Hubbard model
still describes nontrivial correlations among the fermions even in $d=\infty$. This is already apparent from the evaluation of
the second-order diagram in Goldstone perturbation theory
for the correlation energy  at weak coupling \cite{MV89}.
The integral over the three internal momenta (which lead to a nine-dimensional integral in $d=3$) reduces to a single integral in $d= \infty$. Obviously  the calculation is much simpler in $d= \infty$ than in finite dimensions. More importantly, the results for the energy obtained in $d= \infty$ turn out to be very close to those in  $d=3$ (Fig.~\ref{weak-coupling-correlation-energy}) and therefore provide a computationally simple, but quantitatively reliable approximation.

These results showed that microscopic calculations for correlated lattice fermions in $d= \infty$ dimensions were useful and very promising. Further insights followed quickly:
(i) M\"{u}ller-Hartmann  \cite{MH89a} proved that in infinite dimensions only the Hubbard interaction remains dynamical and that the self-energy\index{self-energy} becomes $\bm{k}$-independent, i.e., local in position space, as in the Gutzwiller approximation \cite{MV89,metzner89a} (see footnote \ref{GA-mean-field}), but retains its dynamics\footnote{\label{Selfenergy-Gutz}This result may be understood as follows \cite{Janis91,Janis92a}: The interaction between particles influences their motion. This effect is described by a complex, spatially dependent and dynamical field, the self-energy\index{self-energy} $\Sigma_{\sigma} (\bm{k}, \omega)$. On a lattice with a very large number of nearest neighbors the \emph{spatial} dependence of this field becomes increasingly unimportant and vanished completely in $d = \infty$, as in the mean-field theory of the Ising model. So the field becomes a mean field in position space but retains its full dynamics. In this respect there is a direct analogy to non-interacting electrons in the present of static (``quenched'') disorder,\index{metal-insulator transition} where the self-energy also becomes $\bm{k}$-independent  in the limit $d \to \infty$ (``coherent potential''). The coherent potential approximation  \cite{CPA}\index{coherent potential approximation (CPA)} is a single-site theory where an electron moves through an effective medium described by the self-energy $\Sigma_{\sigma} (\omega)$, and becomes exact in $d = \infty$ \cite{Janis91,vlaming92,Janis92}. In the case of the Hubbard model in the limit $d \to \infty$ the coherent potential is more complicated due to the interaction between the particles (see footnote \ref{CPA-DMFT}).}
\begin{equation}
\Sigma_{\sigma} (\bm{k}, \omega) \stackrel{d \to \infty}{\equiv}
\Sigma_{\sigma} (\omega),
\label{G11.23b}
\end{equation}
whereby typical Fermi liquid features are preserved  \cite{MH89b} (for a discussion see section \ref{Fermi liquid}),
(ii) Schweitzer and Czycholl \cite{Czycholl} demonstrated that  calculations for the periodic Anderson model also become much simpler in high dimensions, and
(iii) Brandt and Mielsch \cite{Brandt} derived the exact solution of the Falicov-Kimball model\index{Falicov-Kimball model} in infinite dimensions by mapping the lattice problem onto a solvable atomic problem in a generalized, time-dependent external field;
they also indicated that in principle such a mapping was even possible for the Hubbard model.\footnote{Alternatively, it can be shown that in the limit  $Z \rightarrow \infty$ the dynamics of the Falicov-Kimball model reduces to that of a non-interacting, tight-binding model on a Bethe lattice with coordination number $Z=3$ which can be solved analytically \cite{vDV90}.}

Due to the $\bm{k}$-independence of the irreducible self-energy the most important obstacle for diagrammatic calculations in
finite dimensions $d \geq 1$, namely the integration over intermediate
momenta, is removed.
At the same time the limit $d \to \infty$ does not affect
the dynamics of the system. Hence, in spite of the
simplifications in position or momentum space, the many-electron problem retains its
full dynamics in $d = \infty$.

\subsubsection{One-particle  propagator and spectral function}

In $d=\infty$ the
one-particle propagator of an interacting lattice fermion system (the ``lattice Green function'') at $T=0$ is then given by
\begin{equation}
G_{\bm{k}, \sigma}^{} (\omega) =
\frac{1}{\omega - \epsilon_{\bm{k}} + \mu - \Sigma_{\sigma} (\omega)}.
\label{G11.32}
\end{equation}
The $\bm{k}$-dependence of $G_{\bm{k}} (\omega)$ now comes
entirely from the energy dispersion
$\epsilon_{\bm{k}}$ of the {\em non}-interacting
particles. This means that in a homogeneous system described by the propagator
\begin{equation}
G_{ij , \sigma} (\omega) = L^{-1} \; \sum_{\bm{k}} G_{\bm{k}, \sigma} (\omega)
e^{i \bm{k} \cdot (\bm{R}_i - \bm{R}_j) }
\label{G11.33}
\end{equation}
its local part, $G_{ii , \sigma} \equiv G_{\sigma} $, is given by
\begin{equation}
G_{ \sigma} (\omega) = L^{-1} \sum\limits_{\bm{k}} G_{\bm{k}, \sigma} (\omega)
= \int\limits_{-\infty}^{\infty}  d E \frac{N_{0} (E)}{ \omega - E + \mu
- \Sigma_{\sigma} (\omega)},
\label{G11.34b}
\end{equation}
where $N_{0} (E)$ is the density of states of the non-interacting system.
 The spectral function of the interacting system (also often called density of states) is given by
\begin{equation}
A_{ \sigma}(\omega) = - \frac{1}{\pi} {\rm Im} G_{ \sigma}(\omega + i0^+).
\label{G11.34c}
\end{equation}

\subsubsection{$\bm{k}$-independence of the self-energy and\index{self-energy} Fermi liquid behavior}\index{Fermi liquid}
\label{Fermi liquid}

The $\bm{k}$-independence of the self-energy allows one to make contact with Fermi liquid theory~\cite{MH89b}.
In general, i.e., even when $\Sigma$ has a $\bm{k}$-dependence, the Fermi surface
is defined by the $\omega = 0$ limit of the denominator of \eqref{G11.32}  (in the paramagnetic phase we can suppress the spin~index)\footnote{We note that the notion of a $(d{-}1)$-dimensional Fermi surface of electrons on $d$-dimensional lattices is no longer meaningful in $d=\infty$ (see footnote \ref{d-infinity-continuum}). Therefore the following discussion addresses the consequences of a momentum independent self-energy for finite-dimensional rather than strictly infinite-dimensional systems.}
\begin{subequations}
\label{G11.40}
\begin{equation}
\epsilon_{\bm{k}} + \Sigma_{\bm{k}} (0) = E_F.
\label{G11.40a}
\end{equation}
According to Luttinger and Ward \cite{LW60} the volume within the
Fermi surface is not changed by interactions, provided the
latter can be treated in perturbation theory.\footnote{Recently necessary and sufficient conditions for the validity of Luttinger's theorem \cite{LW60} based on the Atiyah-Singer index theorem were derived, by which the topological robustness of a generalized Fermi surface may be  quantified  \cite{Bedell2020}.}
This is expressed by
\begin{equation}
n = \sum_{\bm{k} \sigma} \; \Theta \big(E_F - \epsilon_{\bm{k}} - \Sigma_{\bm{k}}(0) \big),
\label{G11.40b}
\end{equation}
where $n$ is the electron density and $\Theta (x)$ is the step function.
The $\bm{k}$-dependence of $\Sigma_{\bm{k}} (0)$
in \eqref{G11.40a}
implies that, in spite of \eqref{G11.40b}, the shape of
the Fermi surface of the interacting system will be quite different from that of the
non-interacting system, except for the rotationally invariant case
$\epsilon_{\bm{k}}=f(|\bm{k}|)$. By contrast, for lattice fermion models in $d = \infty$,
where $\Sigma_{\bm{k}} (\omega) \equiv \Sigma (\omega)$,
the Fermi surface itself, and hence
the enclosed  volume, is not changed by the interaction.
The Fermi energy is simply
shifted uniformly from its non-interacting value
$E_F^0$ to  $E_F = E_F^0{+}\Sigma (0)$, to keep $n$
in \eqref{G11.40b} constant.
Thus $G(0)$, the $\omega  = 0$ value of the
local lattice Green function, and the spectral function
$A(0) = - \frac{1}{\pi}
{\rm Im} G(i0^+)$ are not changed
by the interaction at all. This ``pinning behavior'' is well-known from the single-impurity Anderson model\index{single-impurity Anderson model} \cite{Hewson}. A~renormalization of $N(0)$ can only be due to a
$\bm{k}$-dependence of $\Sigma$.

For $\omega \to 0 $ the self-energy has the property \cite{MH89b}
%%e check change! indicate pricipal value?  {\large P}\hspace{-2.4ex}\int_{-\infty}^\infty
\begin{equation}
{\rm Im} \; \Sigma (\omega) \propto \omega^2,
\label{G11.40c}
\end{equation}
which implies Fermi liquid behavior. The effective mass of the quasiparticles
\begin{equation}
\frac{m^*}{m} = \left. 1 - \frac{d {\rm Re}\Sigma}{d \omega} \right|_{\omega = 0}
%\label{G11.40d}
%\end{equation}
%\begin{equation}
 =  1 + \frac{1}{\pi} \!\int_{-\infty}^{\infty}\! d \omega
\; \frac{{\rm Im} \Sigma (\omega{+}i0^-)}{\omega^2} \geq 1
%\label{G11.40e}
\end{equation}

\end{subequations}
is seen to be enhanced. In particular, the momentum distribution
\begin{equation}
n_{\bm{k}} = \frac{1}{\pi} \!\int_{-\infty}^{0}\! d \omega \; {\rm Im} G_{\bm{k}} (\omega)
\label{G11.41}
\end{equation}
has a discontinuity at the Fermi surface given  by
$n_{k_{F}^-} - n_{k_F^+} = (m^*/m )^{-1}$, where $k_F^\pm = k_F \pm 0^+$.

\subsubsection{Is there a unique $d \to \infty$ limit of the Hubbard model?}

The motivation for the scaling of the hopping amplitude in the limit $d \to \infty$, \eqref{G11.11},  deserves a more detailed discussion. To obtain a physically meaningful mean-field
theory\index{mean-field theory} of a model it is necessary that its internal or free energy  remains finite in the
limit $d$ or $Z\rightarrow \infty $.
While for the Ising model
the scaling $J = {J^*}/Z$, ${J^*}= \text{const.}$, is rather obvious, this is not so for more complicated
models. Namely, fermionic or bosonic many-particle systems are
typically described by a Hamiltonian with
non-commuting terms, e.g., a kinetic energy and an interaction, each of which is associated with a coupling
parameter, usually a hopping amplitude and an interaction, respectively. In such a case
the question of how to scale these parameters has no unique answer
since this depends on the physical effects one wishes to
explore. The scaling should be
performed such that the model remains non-trivial and ``physically interesting'' and that its
energy stays finite in the $d,Z\rightarrow \infty $ limit. Here
``non-trivial'' means that not only $\langle \hat{H}_{0}\rangle $ and
$\langle \hat{H}_{\mathrm{int}}\rangle $ but also their competition, expressed by the commutator $\big\langle \lbrack
\hat{H}_{0},\hat{H}_{\mathrm{int}}]\big\rangle $, should remain finite. In the case of the Hubbard model it would be possible to scale the hopping amplitude as in the mean-field theory of the Ising model, i.e., $t \to t^*/Z, \; t^* = {\rm const.}$,  but then the kinetic energy would be reduced to zero in the limit $d,Z \to \infty$, making the resulting model uninteresting (but not unphysical) for most purposes. For the bosonic Hubbard model the situation is even more subtle, since the kinetic energy has to be scaled differently depending on whether it describes the normal or the Bose-Einstein condensed fraction; for a discussion see ref.~\cite{B-DMFT}. Hence, in the case of a many-body system described by a Hamiltonian with several terms, a mean-field solution in the limit $d\rightarrow \infty $ depends crucially on how the scaling of the model parameters is chosen.

\section[Dynamical mean-field theory (DMFT) of correlated electrons and its applications]{Dynamical mean-field theory (DMFT)\newline of correlated electrons and its applications\index{dynamical mean-field theory (DMFT)}}\index{electronic correlations}

The diagrammatic simplifications of quantum many-body theory in infinite spatial dimensions provide the basis for the construction of a comprehensive mean-field theory of lattice fermions
 which is diagrammatically controlled and whose free energy has no unphysical singularities. The construction is based on the scaled Hamiltonian \eqref{scaled-HM}.
The self-energy  is then momentum independent, but retains its frequency dependence and thereby describes the full many-body dynamics of the interacting system.\footnote{This is in contrast to Hartree(-Fock) theory where the self-energy acts only as a static potential.}
The resulting theory is both mean-field-like and dynamical and therefore represents a \emph{dynamical} mean-field theory (DMFT) for lattice fermions which is able to describe genuine correlation effects as will be discussed next.

\subsection{The self-consistent DMFT equations}\index{self-consistent equation}
\label{DMFT-equations}

\begin{figure}
\centerline{\includegraphics[width=0.9\textwidth]{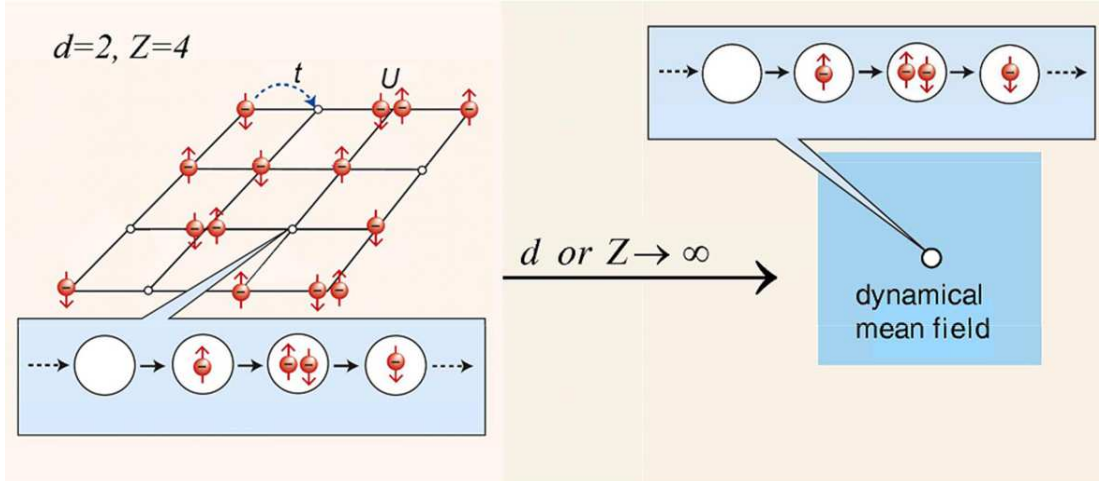}}
\caption{In the limit $d$ or $Z\rightarrow \infty $ the Hubbard model\index{Hubbard model} effectively reduces to a dynamical single-site problem which may be viewed as a lattice site embedded in a $\bm{k}$-independent, dynamical fermionic mean field. Electrons can hop from the mean field onto this site and back, and interact on the site as in the original Hubbard model (see Fig.~\ref{Hubbard_model}).
The local dynamics of the electrons is independent of the dimension or coordination number and therefore remains unchanged.}
\label{DMFT}
\end{figure}

In $d=\infty$, lattice fermion models with a local interaction effectively reduce to a
%an effective many-body problem whose dynamics corresponds to that of correlated fermions on a
single site embedded in a dynamical mean field provided by the other interacting fermions as illustrated in Fig.~\ref{DMFT}. The self-consistent DMFT equations can be derived in different ways depending on the physical interpretation of the correlation problem emerging in the limit $d,Z \to \infty$ \cite{Janis91,Georges92,Jarrell92}; for a discussion see ref.~\cite{Vollhardt-Salerno}.
The mapping of the lattice electron problem onto a single-impurity Anderson model\footnote{The mapping itself can be performed without approximation, but leads to a complicated coupling between the impurity and the bath which makes the problem intractable. This can be solved in the limit $d\rightarrow \infty $ when the momentum dependence of the self-energy\index{self-energy} drops out.}\index{single-impurity Anderson model} with a self-consistency condition in $d=\infty$ introduced by Georges and Kotliar \cite{Georges92}, which was also employed by Jarrell \cite{Jarrell92},
turned out to be the most useful approach, since it made a connection with the well-studied theory of quantum impurities \cite{Hewson}, for whose solution efficient numerical codes such as the quantum Monte-Carlo (QMC) method \cite{Hirsch86}  had already been developed and were readily available.\footnote{\label{CPA-DMFT}Alternatively Jani\v{s} derived the self-consistent DMFT equations by generalizing the coherent potential approximation (CPA)\index{coherent potential approximation (CPA)} \cite{Janis91}. In the CPA quenched disorder\index{metal-insulator transition} acting on non-interacting electrons is averaged and produces a mean field, the ``coherent potential''. In the case of the Hubbard model in $d=\infty$ the
fluctuations generated by the Hubbard interaction may be treated as ``annealed'' disorder acting on non-interacting electrons \cite{Janis92a} which, after averaging, produce a mean field, the self-energy.\index{self-energy} Numerical solutions of the DMFT equations starting from the CPA point of view have not yet been developed.}
For a detailed discussion of the foundations of DMFT see the review by Georges,  Kotliar, Krauth, and Rozenberg \cite{georges96} and the lecture by Kollar at the J{\"u}lich Autumn School 2018 \cite{Julich-2018}; an introductory presentation can be found in ref.~\cite{dmft_phys_today}.

For $T>0$ the self-consistent DMFT equations are given by:\\
(I) the \emph{local propagator} $G_{\sigma }(i\omega _{n})$, which is expressed by a functional integral as
\begin{equation}
G_{\sigma }(i\omega _{n})=-\frac{1}{{\mathcal{Z}}}\int \prod_{\sigma
}Dc_{\sigma }^{\ast }Dc_{\sigma }[c_{\sigma }(i\omega _{n})\,c_{\sigma }^{\ast
}(i\omega _{n})]\,\exp\!\big(S_{\mathrm{loc}}\big)
\label{Vollhardt:impurity_problem}
\end{equation}
with the partition function
\begin{equation}
{\mathcal{Z}}=\int \prod_{\sigma }Dc_{\sigma }^{\ast }Dc_{\sigma }\exp\!\big(-S_{\mathrm{loc}}\big)
\end{equation}
and the local action
\begin{equation}
S_{\mathrm{loc}}=- \!\int_0^{\beta}\!\! d\tau_1 \!\int_0^{\beta}\!\! d\tau_2
\sum_{\sigma} c^{*}_{\sigma} (\tau_1)\, \mathcal{G}^{-1}_{\sigma}(\tau_1{-}\tau_2)\, c _{\sigma}(\tau_2)
+ U\!\int_0^{\beta}\!\! d\tau\,
c^{*}_{\uparrow}(\tau)c_{\uparrow}(\tau)c^{*}_{\downarrow}(\tau)c_{\downarrow}(\tau).
\label{Vollhardt:local_action}
\end{equation}
Here $\mathcal{G}_{\sigma}$ is the effective local propagator (also called ``bath Green function'', or ``Weiss mean field''\footnote{This expresses the fact that $\mathcal{G}$ describes the coupling of a single site to the rest of the system, similar to the Weiss mean-field $h_{\rm MF}$ in the mean-field theory of the Ising model (see section \ref{sec:ising}). However, in the case of the DMFT the mean field depends on the frequency, i.e., is dynamical. It should be noted that, in principle, both local functions $\mathcal{G}_{\sigma
}(i\omega _{n})$ and $\Sigma _{\sigma }(i\omega _{n})$
can be viewed as a dynamical
mean field\index{mean field} since both
enter in the bilinear term of the local action (\ref{Vollhardt:local_action}).}),
which is defined by a Dyson equation
\begin{eqnarray}
\mathcal{G}_{\sigma}(i\omega _{n}) = \Big( \big(G_{\sigma }(i\omega _{n})\big)^{-1} + \Sigma _{\sigma }(i\omega _{n})\Big)^{-1}.
\label{bath}
\end{eqnarray}

Furthermore, by identifying the
local propagator (\ref{Vollhardt:impurity_problem}) with the Hilbert transform of the lattice Green function
\begin{equation}
G_{\bm{k}\,\sigma }(i\omega _{n})=\frac{1}{i\omega _{n}-\epsilon _{\bm{k}}+\mu -\Sigma _{\sigma }(i\omega _{n})},
\end{equation}
(this identification is exact in $d=\infty $ \cite{georges96}), one obtains \\
(II) the \emph{self-consistency condition}
\begin{eqnarray}
G_{\sigma }(i\omega _{n}) = \frac{1}{L} \sum_{\bm{k}}G_{\bm{k}\,\sigma }(i\omega_{n}) &=&
\underset{-\infty }{\overset{\infty }{\int }}d\epsilon \frac{N(\omega )}{i\omega _{n}-\epsilon +\mu -\Sigma _{\sigma }(i\omega _{n})}
\label{Vollhardt:local_prop-1}\\[1ex]
&=& G_{\sigma }^{0}\big(i\omega _{n}{-}\Sigma_{\sigma }(i\omega _{n})\big).
\label{Vollhardt:local_prop}
\end{eqnarray}
In (\ref{Vollhardt:local_prop-1}) the ionic lattice enters only through the density of states of the non-interacting electrons.\footnote{Eq.~(\ref{Vollhardt:local_prop}) illustrates the mean-field nature of the DMFT equations  very clearly: the local Green function  of the interacting system is given by the non-interacting Green function $G_{\sigma }^{0}$ at the renormalized energy $i\omega _{n}- \Sigma _{\sigma }(i\omega _{n})$, which corresponds to the energy measured relative to the energy $\Sigma _{\sigma }(i\omega _{n})$\index{self-energy} of the surrounding fermionic bath, the dynamical mean field.\index{mean field}}

The self-consistent DMFT equations\index{self-consistent equation} can be solved iteratively: starting with an initial guess for the self-energy $\Sigma _{\sigma }(i\omega _{n})$ one obtains the local propagator  $G_{\sigma }(i\omega _{n})$ from \eqref{Vollhardt:local_prop-1} and thereby the bath Green function $\mathcal{G}_{\sigma
}(i\omega _{n})$ from \eqref{bath}. This determines the local action \eqref{Vollhardt:local_action} which is needed to compute
a new value for the local propagator $G_{\sigma }(i\omega _{n})$ from (\ref{Vollhardt:impurity_problem}). By employing the old self-energy a new bath Green function $\mathcal{G}_{\sigma}$ is calculated and so on, until convergence is reached.

It should be stressed that although the DMFT  corresponds to an effectively local problem, the propagator $G_{\bm{k}}(\omega)$  depends on the crystal momentum $\bm{k}$ through the dispersion relation $\epsilon_{\bm{k}}$ of the non-interacting electrons. But there is no \emph{additional} momentum-dependence through the self-energy, since this quantity is local within the DMFT.

Solutions of the self-consistent DMFT  equations require the extensive application of numerical methods, in particular quantum Monte-Carlo (QMC) simulations \cite{Jarrell92,georges96} with continuous-time QMC \cite{CT-QMC2011} still as the method of choice, the numerical renormalization group \cite{NRG-RMP}, the density matrix renormalization group \cite{DMRG}, exact diagonalization \cite{georges96}, Lanczos procedures \cite{ED+Lanczos}, and solvers based on matrix product states \cite{MPS}  or tensor networks \cite{Tensor-networks}. Here the recent development of impurity solvers making use of machine learning opens new perspectives \cite{ML}.

\subsubsection{Characteristic features of DMFT}

In DMFT the mean field is dynamical, whereby local quantum
fluctuations are fully taken into account, but is local (i.e., spatially independent) because of the infinitely many neighbors of every lattice site (``single-site DMFT'').
The only approximation of DMFT when applied in $d<\infty$ is the neglect of the $\bm{k}$-dependence of the self-energy.\index{self-energy} DMFT
provides a comprehensive, non-perturbative, thermodynamically consistent and diagrammatically controlled approximation scheme for the investigation of correlated lattice models at all interaction strengths, densities, and temperatures \cite{georges96,dmft_phys_today}, which can resolve even low energy scales.
It describes fluctuating moments, the renormalization of quasiparticles and their damping, and is especially valuable for the study  of correlation problems at intermediate couplings, where no other investigation methods are available.
Unless a symmetry is broken the $\bm{k}$-independence of the self-energy implies typical Fermi-liquid properties of the DMFT solution.

Most importantly, with DMFT it is possible to compute electronic correlation effects quantitatively in such a way that they can be tested experimentally, for example, by electron spectroscopies. Namely, DMFT describes the correlation induced transfer of spectral weight and the finite lifetime of quasiparticles through the real and imaginary part of the self-energy, respectively.
This greatly helps to understand and characterize the Mott metal-insulator transition (MIT) to be discussed next.

\subsection{Mott transition, ferromagnetism, and topological properties}\index{Mott transition}\index{topological properties}

Intensive theoretical investigations of the Hubbard model and related correlation models using DMFT over the last three decades have  provided a wealth of new insights into the physics described by this fundamental fermionic interaction model.
In the following subsection only a few exemplary results will be discussed; more detailed presentations can be found in refs.~\cite{georges96,Vollhardt-Salerno}, and the lecture notes of the J\"{u}lich Autumn Schools in 2011, 2014, and 2018 \cite{Juelich-lecture-notes}.

\subsubsection{Metal-insulator transitions}\index{metal-insulator transition}
\label{sec:mit}

\paragraph{Mott-Hubbard transition}

The interaction-driven transition between a paramagnetic metal and a
paramagnetic insulator, first discussed by Mott \cite{mott} and referred to as ``Mott metal-insulator transition'' (MIT), or ``Mott-Hubbard MIT'' when studied within the Hubbard model, is  one of the most intriguing phenomena in
condensed matter physics~\cite{Mott90,TokuraRMP}.
This transition is
a consequence of the quantum-mechanical competition
between the kinetic energy of the electrons and their interaction $U$\hspace{-0.15ex}: the kinetic energy prefers the electrons to be mobile (a wave effect) which invariably leads to their interaction (a particle effect). For large values of $U$ doubly occupied sites become energetically too costly. The system can reduce its total energy by localizing the electrons, which leads to a MIT.
Here the DMFT provided detailed insights into the nature of the Mott-Hubbard-MIT  for all values of the interaction $U$ and temperature $T$ \cite{georges96,MIT,dmft_phys_today,Vollhardt-Salerno}. A microscopic investigation of the Mott MIT obtained within DMFT from a Fermi-liquid point\index{Fermi liquid} of view was performed only recently \cite{Krien2019}.

While at small $U$ the interacting system can be described by coherent quasiparticles
whose spectral function (``density of states'' (DOS)) still resembles that of the free electrons, the DOS in the Mott
insulating state consists of two separate incoherent ``Hubbard
bands'' whose centers are separated approximately by the energy
$U$ (here we discuss only the half-filled case without magnetic order).
At intermediate values of $U$ the spectrum then has a
characteristic three-peak structure which is qualitatively similar to that of the single-impurity
Anderson model \cite{SIAM} and which
is a consequence of
the three possible occupations of a lattice site: empty, singly occupied (up or down), and doubly occupied.

At $T=0$ the width of the quasiparticle peak vanishes at a critical value of $U$ which is of the order of the band width.
So the Mott-Hubbard MIT occurs at intermediate coupling and therefore belongs to the hard problems in many-body theory,
where most analytic approaches fail and investigations have to rely on numerical methods. Therefore several features of the Mott-Hubbard MIT near the transition point are still not sufficiently understood even within DMFT.
Actually, it was recently argued that the
Mott-Hubbard transition in the infinite-dimensional one-band Hubbard
model may be understood as a topological phase transition,\index{topological properties} where the insulating state is the topological phase, and the transition from the metallic (Fermi liquid) to the insulating state involves domain wall dissociation \cite{MIT-topological2020}.

At $T>0$ the Mott-Hubbard MIT  is found to be first order
and is associated with a hysteresis region in
the interaction range $U_{\rm c1}<U<U_{\rm c2}$ where $U_{\rm c
1}$ and $U_{\rm c 2}$ are the values at which the insulating and
metallic solution, respectively, vanish
\cite{georges96,MIT}; for more detailed discussions see refs.\cite{georges96,dmft_phys_today,Vollhardt-Salerno}.
The hysteresis region
terminates at a critical point,
above which
the transition becomes a smooth crossover from
a ``bad metal'' to a ``bad insulator''; for a schematic plot of the phase diagram see fig.~3 of ref.~\cite{dmft_phys_today}.
Transport in the incoherent region above the critical point shows remarkably rich properties, including scaling behavior
\cite{Transport-MIT}.

 Mott-Hubbard MITs are found, for example, in transition metal oxides  with
partially filled bands. For such systems  band theory
typically predicts metallic behavior. One of the most famous  examples is V$_{2}$O$_{3}$ doped with Ti or Cr~\cite{V2O3}.
However, it is now known that certain organic materials
are better realizations of the single-band Hubbard model
without magnetic order and
allow for much more controlled investigations of the Mott state and the Mott MIT~\cite{MIT-in-organics}.

\paragraph{Metal-insulator transitions in the presence of disorder}\index{metal-insulator transition}\index{metal-insulator transition}
\label{Electronic Correlations and Disorder}

DMFT also provides a theoretical framework for the investigation of correlated electrons in the presence of disorder.
When the effect of local disorder is taken into account
through the arithmetic mean of the local DOS (LDOS) one obtains, in the
absence of interactions, the coherent potential approximation (CPA)\index{coherent potential approximation (CPA)}
\cite{vlaming92,Janis92}; for a discussion see ref.~\cite{Vollhardt-Salerno}. However, CPA cannot describe Anderson
localization. To overcome this deficiency a variant of the DMFT was formulated where the
\emph{geometrically} averaged LDOS is computed from the solutions of the
self-consistent stochastic DMFT equations and is then fed into the self-consistency cycle \cite{Dobrosavljevic97+03+15}.
This corresponds to a ``typical medium theory'' which can describe the Anderson transition of non-interacting electrons.
By implementing this scheme into DMFT to study the properties of disordered electrons in the presence of
interactions it is possible to compute the phase diagram of the
Anderson-Hubbard model   \cite{Byczuk05+Byczuk10}.

\subsubsection{Band ferromagnetism}\index{ferromagnetism}
\label{Band-ferromagnetism}

The Hubbard model was introduced in 1963 \cite{Gutzwiller,HubbardI,Kanamori} in an attempt to explain metallic (``band'') ferromagnetism in 3$d$ metals such as Fe, Co, and Ni starting from a microscopic point of view. However, at that time investigations of the model employed strong and uncontrolled approximations. Therefore  it was uncertain for a long time whether the Hubbard model can explain band ferromagnetism at realistic temperatures, electron densities, and interaction strengths in $d>1$ at all.
Three decades later it was shown with DMFT that  on generalized fcc-type lattices  in $d=\infty$ (i.e., on ``frustrated'' lattices with large spectral weight at the lower band edge) the Hubbard model indeed describes metallic ferromagnetic phases  in large regions of the phase diagram \cite{FM-Ulmke,ferromagnetism}.
In the paramagnetic phase the
susceptibility $\chi _{F}$ obeys a Curie-Weiss law \cite{Byczuk2002}, where the Curie temperature $T_{C}$  is now much lower than that obtained by Stoner theory, due to many-body effects. In the ferromagnetic phase the magnetization $M$
%(which vanishes at the same $T_{C}$)
is consistent with a Brillouin function as originally derived for localized spins, even for a \emph{non-integer}
magneton number  as in 3$d$
transition metals.
Therefore, DMFT accounts for the behavior of both the magnetization and the susceptibility of
band ferromagnets \cite{FM-Ulmke,ferromagnetism}; see section 6.1 of my lecture notes at the 2018 J\"{u}lich Autumn School  \cite{Julich-2018}. As the Mott MIT, band ferromagnetism is a hard intermediate-coupling problem.

\vspace{-1ex}
\subsubsection{Topological properties of correlated electron systems}\index{topological properties}
\vspace{-1ex}

The interest in non-trivial topological properties of electronic systems sparked by the theory of the quantum Hall effect greatly increased when it was realized that the spin-orbit interaction can generate topologically insulating behavior \cite{Topological-RMP}. Initially, investigations focused on topological features of non-interacting systems. But during the last decade the influence of electronic interactions on these topological properties received more and more attention. Here DMFT turned out to be a useful tool. For example, DMFT was employed to study interaction effects in two-dimensional topological insulators \cite{Topological-Kawakami}, and to analyze the robustness of the Chern number in the Haldane-Hubbard model \cite{Topological-Valenti} as well as of the topological quantization of the Hall conductivity of correlated electrons at $T>0$ \cite{Topological-Rubtsov}. Furthermore, to better understand the topological phase transition from a Weyl-semimetal to a Mott insulator the topological properties of quasiparticle bands were computed \cite{Topological-Hofstetter}. DMFT also made it possible to explore topological phase transitions in the Kitaev model in a magnetic field and to calculate the corresponding phase diagrams \cite{Topological-Liang}. Correlation-induced topological effects can even arise from non-Hermitian properties of the
single-particle spectrum in equilibrium systems \cite{Topological-Kawakami2}.
Recently it was demonstrated explicitly that a topologically nontrivial multiorbital Hubbard model remains well-defined and nontrivial in the limit  $d \to \infty$ for arbitrary, but even, $d$ \cite{Topological-Potthoff}.

\vspace{-1ex}
\subsubsection{Nonequilibrium DMFT}\index{nonequilibrium DMFT}
\vspace{-1ex}

The study of correlated electrons out of equilibrium by employing a generalization of the DMFT  has become  another fascinating new research area \cite{RMP-Noneq-DMFT2014}. Nonequilibrium DMFT is able to explain, for example, the results of time-resolved  electron spectroscopy experiments, where femtosecond pulses are now available in a wide frequency range. In such experiments a probe is excited and the subsequent relaxation is studied. One recent example is the ultrafast dynamics of doubly occupied sites in the photo-excited quasi-2$d$ transition-metal dichalcogenide 1\emph{T}-TaS$_2$ \cite{1T-TaS2}.
Such excitations may even result in long-lived, metastable (``hidden'') states \cite{hidden}.

\vspace{-1.5ex}
\subsection{Correlated electrons in bulk materials, surfaces, and nanostructures}\index{electronic correlations}
\label{LDA+DMFT}
\vspace{-1ex}

The development and application of theoretical techniques to explore the basic features of the physics described by the one-band Hubbard model took several decades.
During that time first-principles investigations of the even more complicated many-body problem of correlated \emph{materials} were clearly out of reach.
The electronic properties of solids were  mainly studied within density-functional theory (DFT) \cite{DFT1-DFT2},
e.g., in the local density approximation (LDA) \cite{LDA}, the generalized gradient approximation (GGA) \cite{PB96}, and the LDA+U method \cite{Anisimov91}.
Those approaches can describe the ground state properties of many simple elements and semiconductors, and even of some insulators, quite accurately, and often predict the magnetic and orbital properties \cite{LDA} as well as the crystal structures of many solids correctly~\cite{Baroni01}.
However, these methods fail to describe the electronic
and structural properties of correlated paramagnetic materials since they miss characteristic features of correlated electron systems such as heavy quasiparticle behavior and Mott physics.
This situation changed dramatically with the advent of DMFT.

\vspace{-1.5ex}
\subsubsection{DFT+DMFT and $GW$+DMFT}\index{DFT+DMFT}\index{GW+DMFT}
\vspace{-1ex}

The computational scheme introduced by Anisimov \emph{et al.} \cite{Anisimov97} and Lichtenstein and Katsnelson \cite{Lichtenstein98}, which merges material-specific DFT-based approximations with the many-body DMFT and which is now denoted by DFT+DMFT (or rather more specifically LDA+DMFT, GGA+DMFT, etc.),
provides a powerful new method for the microscopic computation
of the electronic, magnetic, and structural properties of correlated materials
from first principles even at
finite temperatures \cite{LDA+DMFT-Held,LDA+DMFT-Kotliar06,LDA+DMFT-Katsnelson08,LDA+DMFT-FOR1346-Report-2017,Vollhardt2020}. In particular, this approach naturally accounts for the existence of local moments in the paramagnetic phase. By construction, DFT+DMFT includes the correct quasiparticle physics and the corresponding energetics, and reproduces the DFT results in the limit of weak
Coulomb interaction $U$\!.\, Most importantly, DFT+DMFT describes the
correlation-induced many-body dynamics of strongly correlated electron materials at all values of the Coulomb interaction and doping.

As in the case of the single-band Hubbard model the many-body model of correlated materials constructed within the DFT+DMFT scheme consists of two parts: an effective kinetic energy obtained by DFT which describes the material-specific band structure of the uncorrelated electrons, and the local interactions between the electrons in the same orbital as well as in different orbitals. Here the static contribution of the electronic interactions already included in the DFT-approximations must be subtracted to avoid double counting  \cite{Anisimov97,Lichtenstein98,LDA+DMFT-Held,LDA+DMFT-Kotliar06,LDA+DMFT-Katsnelson08,LDA+DMFT-FOR1346-Report-2017}. Such a correction is not necessary in the fully diagrammatic, but computationally very demanding $GW$+DMFT approach, where the LDA/GGA input is replaced by the $GW$ approximation \cite{GW+DMFT}. The resulting many-particle problem with its numerous energy bands and local interactions is then solved within DMFT, typically by CT-QMC.

\paragraph{Bulk materials}
The application of DFT+DMFT made investigations of correlated materials much more realistic and led to the discovery of novel physical mechanisms and correlation phenomena. Take, for example, the Mott MIT.\index{Mott transition} Within the single-band Hubbard model the Mott MIT was originally explained as a transition where the effective mass of quasiparticles diverges (``Brinkman-Rice scenario'') \cite{BR70}. After DMFT had opened the way to study multi-band models, an orbitally-selective Mott MIT was identified \cite{Georges+Biermann}. Then, with the advent of DFT+DMFT, a site-selective Mott MIT was discovered in Fe$_{2}$O$_{3}$ with its numerous energy bands and local interactions \cite{Leonov+Abrikosov}.
With DFT+DMFT it was also shown that the non-trivial topological properties of $\alpha$-Ce are driven by the $f$-$d$ band inversion, which originates from the formation of a coherent $4f$ band around the Fermi energy \cite{Topological-Min}.

DFT+DMFT has been remarkably successful in the investigation
of correlated materials, including transition metals and their oxides, manganites, fullerenes, Bechgaard salts, $f$-electron materials, magnetic superconductors,  and Heusler alloys  \cite{LDA+DMFT-Held,LDA+DMFT-Kotliar06,LDA+DMFT-Katsnelson08,LDA+DMFT-FOR1346-Report-2017}.
  In particular, the study of Fe-based pnictides and chalcogenides led to the new insight that in  metallic multi-orbital materials the intra-atomic exchange  $J$  can also induce strong correlations \cite{Hunds-metal}. Clearly DMFT-based approaches will be very useful for the future design of correlated materials \cite{Adler-Kotliar-2018}, such as materials with a high thermopower for thermoelectric devices
which can convert waste heat into electric energy \cite{thermo}.

\paragraph{Surfaces, layers, and nanostructures}
DMFT studies of inhomogeneous systems greatly improved our understanding of correlation effects at surfaces and interfaces, in thin films and multi-layered nanostructures \cite{Heterostructures-etc} and, most recently, in infinite layer nickelates \cite{Nickelates2022}. Thereby they provided a new understanding of potential functionalities of such structures and their application in electronic devices.
DMFT has been extended to study correlations also in finite systems such as nanoscopic
 conductors and molecules \cite{molecules}. In this way many-body effects were shown to be important even in biological matter, e.g., in the kernel of hemoglobin and molecules with important biological functions \cite{Weber}.

\section{Beyond mean-field theory}\index{corrections to mean-field theory}

Mean-field approximations provide useful information on the general physical properties of many-body problems.
In particular, DMFT with its dynamical but local self-energy has been a breakthrough for the investigation and explanation of electronic  correlation effects in models and materials.
Although it is an approximation when used in  $d<\infty$, experiments with cold atoms in optical lattices\index{cold atoms in optical lattices} demonstrated that  DMFT can be remarkably accurate in $d=3$ \cite{Mott2}. A dynamical, local self-energy was also shown to be well justified in iron pnictides and chalcogenides \cite{local1} as well as in Sr$_{2}$RuO$_{4}$ \cite{local2}.
Nevertheless mean-field results clearly cannot explain correlation phenomena occurring on finite length scales, the critical behavior at thermal or quantum phase transitions, or unconventional superconductivity observed in finite-dimensional systems. In such cases corrections to mean-field theory must be included.

\subsection{\texorpdfstring{$1/d$}{1/d} corrections}
\label{1-over-d-corrections}

For mean-field theories\index{mean-field theory} derived in the limit $d \to \infty$ corrections can be obtained, in principle, by performing an expansion in the parameter $1/d$ around the mean-field results:

\paragraph{Mean-field theory of the Ising model}\index{Ising model}
$1/d$-expansions  \cite{Fisher+Gaunt} around this paradigmatic mean-field theory\footnote{For the ferromagnetic Ising model high-temperature expansions can be converted to $1/d$-expansions around mean-field theory \cite{Georges+Yedidia91}.} have been performed in great detail, resulting in high-order expansions of the free energy and of susceptibilities \cite{Butera2012}. Expansions in $1/d$ to determine the critical temperature $T_C$ in $d < \infty$ appear to be asymptotic \cite{Fisher+Singh}.

\paragraph{Gutzwiller approximation}\index{Gutzwiller wave function}
Systematic $1/d$-expansions around the Gutzwiller approximation lead to excellent agreement with results obtained by variational Monte-Carlo techniques in $d=2,3$ \cite{vanDongen89,metzner89a,Gebhard90}. However, finite orders in $1/d$ do not remove the Brinkman-Rice transition (section \ref{Brinkman-Rice}) found in $d = \infty$.

\paragraph{Coherent potential approximation (CPA)}\index{coherent potential approximation (CPA)}
The CPA corresponds to the exact solution of the Anderson disorder model in $d = \infty$ \cite{vlaming92,Janis92}, but does not describe Anderson localization. Corrections to the CPA self-energy in $1/d$ can be calculated, but bare perturbation theory leads to unphysical divergencies at the band edges \cite{vlaming92}. Indeed, such expansions do not fulfill the condition ${\rm Im} \big(z{-}\Sigma (z)\big) \lessgtr 0 $ for $ {\rm Im} z \lessgtr 0$ (Herglotz analyticity) which would guarantee the non-negativity of the density of states. This also holds for $1/d$-corrections to the conductivity of disordered electrons. The conductivity has the correct analytic properties only if the self-energy is properly connected to the irreducible vertex function by a Ward identity \cite{Janis99,Janis-DV01}. To this end vertex corrections need to be included which, however, mix corrections with different powers in $1/d$. Self-consistent $1/d$-expansions around CPA which are able to describe the Anderson localization transition are presently not in sight. Here numerical cluster approaches to be discussed in section \ref{DMFT-cluster} are more successful.

\paragraph{DMFT of correlated lattice electrons}\index{dynamical mean-field theory (DMFT)}
Systematic calculations of $1/d$-corrections to the DMFT start with a Luttinger-Ward functional of the non-local Green function $G_{ij , \sigma} (i\omega_n)$, from which the non-local self-energy $\Sigma_{ij , \sigma} (i\omega_n)$ is obtained by functional derivative. To calculate first-order corrections in $1/d$ one needs to consider only one pair of nearest-neighbor sites \cite{Schiller95}. This generalizes the single-impurity problem of the DMFT to a two-impurity problem. It is then possible, in principle, to formulate a self-consistent, thermodynamically consistent approximation which is correct to order $1/d$ \cite{Schiller95,Janis99}. This scheme requires an exact cancellation of certain diagrams in the approximation. Unfortunately, numerical computations within this approach are unstable and can easily lead to acausal solutions. Although the scheme was modified such that the diagrammatic cancellation is assured at each iteration step \cite{Zarand00} and thereby provided a causal solutions in test calculations for the Hubbard model, acausal behavior cannot be ruled out in general. Therefore it is still not clear whether, and how, controlled and thermodynamically consistent $1/d$ expansions around DMFT can be constructed. Apparently analytic calculations in this directions were not further pursued in view of the successes of numerical cluster approaches
which, although not systematic in $1/d$, are explicitly causal \cite{Cluster} (see below).

\subsection{Beyond DMFT}\index{dynamical mean-field theory (DMFT)}\index{cluster extensions of DMFT}
\label{DMFT-cluster}

There are different, mostly  numerical, techniques to include non-local correlations into the DMFT; for a review see ref.~\cite{nonlocal}; in the following we mention four approaches:

\paragraph{Extended DMFT}
This is an early strategy to include intersite quantum fluctuations into DMFT  \cite{EDMFT}, where the interaction strength of a nearest-neighbor density-density interaction is scaled such that its fluctuation part contributes even in the large $d$ limit.

\paragraph{Cluster extensions}
The dynamical cluster approximation (DCA) \cite{DCA} and the cellular DMFT (CDMFT) \cite{CDMFT} map a lattice model onto a cluster of sites, which is then self-consistently embedded in a dynamical mean field. Thereby it has become possible to compute, for example, typical features of unconventional superconductivity in the Hubbard model in $d=2$, such as the interplay of antiferromagnetism and $d$-wave pairing as well as pseudogap behavior  \cite{Cluster}, but also signatures of Anderson localization in disordered systems \cite{Cluster-Anderson-localization}.

\paragraph{Diagrammatic generalizations}
By extending the DMFT on a diagrammatic level through the inclusion of non-local contributions, corrections to the local self-energy of the DMFT can be calculated explicitly. Here the dynamical vertex approximation\index{dynamical vertex approximation} (D$\mathrm{\Gamma}$\hspace{-.18ex}A) \cite{DGA} and the dual fermion theory \cite{DF} are powerful approaches. For example, they provided new insights into the mechanism of superconductivity arising from purely repulsive interactions, e.g., in the two-dimensional Kondo lattice model \cite{DMFT-superconductivity-Otsuki}  and the Hubbard model \cite{DMFT-superconductivity-Held}. In particular, in the repulsive Hubbard model a specific set of local particle-particle diagrams was identified which describe a strong screening of the bare interaction at low frequencies. Thereby antiferromagnetic spin fluctuations are suppressed, which in turn reduce the pairing interaction. Thus dynamical vertex corrections were found to reduce  $T_c$ strongly \cite{DMFT-superconductivity-Held}. With these approaches one can also determine critical behavior,  not only in the vicinity of thermal phase transitions ($T>0$) \cite{criticality1} but also near quantum phase transitions ($T=0$) \cite{criticality2}.

\paragraph{Functional renormalization group (fRG)}\index{functional renormalization group (fRG)}
In this approach the fRG flow \cite{fRG} does not start from the bare action of the system, but rather from the DMFT solution \cite{DMFT+fRG}. Local correlations are thus included from the beginning, and nonlocal correlations are generated by the fRG flow, as  demonstrated for the two-dimensional Hubbard model
\cite{DMFT+fRG}.
\clearpage

\section{Summary}

In the absence of exact solutions of models in statistical and condensed matter physics in dimensions $d{\,=\,}2$ and $3$, mean-field solutions give indispensable, albeit approximate, information about the properties of these models. A well-established strategy for deriving a mean-field theory is the solution of a model in infinite dimensions or on a lattice with infinite coordination.
In fact, solutions obtained in this way are sometimes referred to as ``mean-field solution'', even if the physical meaning of the ``mean field'' is not directly apparent.

In particular, the exact solution of the Hubbard model in $d=\infty$, which corresponds to a dynamical mean-field theory (DMFT) of correlated lattice fermions, has now become the generic mean-field theory of correlated electrons. It provides a comprehensive, non-perturbative and thermodynamically consistent approximation scheme for the investigation of correlated fermi\-ons, especially electrons in solids and cold fermionic atoms in optical lattices, in finite dimensions. Non-local extensions of the DMFT  make it possible to explore and explain correlation effects which occur on the scale of several lattice constants and at thermal and quantum phase transitions.
Furthermore, the combination of DMFT with methods for the computation of electronic band structures
resulted in a conceptually new theoretical framework for the realistic study of correlated materials.

The further development of this approach and its applications is a subject of active research.

\subsection*{Acknowledgments} I thank Krzysztof Byczuk,  V\'{a}clav Jani\v{s}, Marcus Kollar, and Walter Metzner for discussions and useful comments.

%%%%%%%%%%%%%%% References %%%%%%%%%%%%%%%%%

%% or, should you use bibtex:
%\bibliographystyle{correl}
%\bibliography{exa}

\clearchapter

\end{document}